%% file: main.tex
\documentclass[10pt,twocolumn]{research-article-2col}

\templatetype{research-article-2col} 

\usepackage{rotating}  
\usepackage{booktabs}
\usepackage{tablefootnote}
\usepackage{xcolor}
\usepackage{pdfpages}

\usepackage{hyperref}
\hypersetup{
    colorlinks=true,
    linkcolor=blue,
    filecolor=magenta,      
    urlcolor=blue,
    }   
\input{abbreviations}

\title{Multinuclear fingerprinting}

\author[1,2]{Gonzalo G. Rodriguez}
\author[1,3,4]{Zidan Yu} 
\author[1,5]{Lauren O'Donnell} 
\author[1,6]{Martijn A. Cloos} 
\author[1,3*]{Guillaume Madelin} 

\affil[1]{Center for Biomedical Imaging, Department of Radiology, New York University Grossman School of Medicine, New York, NY, USA}
\affil[2]{NMR Signal Enhancement, Max Planck Institute for Multidisciplinary Sciences, G\"ottingen, Germany}
\affil[3]{Vilcek Institute of Graduate Biomedical Sciences, New York University Grossman School of Medicine, New York, NY, USA}
\affil[4]{Philips, Best, The Netherlands}
\affil[5]{JEOL USA, Peabody, MA, USA}
\affil[6]{Donders Center for Cognitive Neuroimaging, Radboud University, Nijmegen, The Netherlands}
\affil[*]{Corresponding author: guillaume.madelin@nyulangone.org}

\begin{abstract}

We developed a new magnetic resonance imaging method called multinuclear fingerprinting (MNF) which leverages simultaneously-acquired proton (\hhx) and sodium (\nax) data to generate seven quantitative maps of the whole brain: proton density (PD), \tone and \ttwo relaxation times from water, and tissue sodium concentration (TSC), \tonex, \ttwoshort and \ttwolong from \naplus ions. MNF consists of two parts: (1) simultaneous \hhna magnetic resonance fingerprinting (MRF), and (2) a super-resolution (SR) algorithm to increase the \na resolution to match the \hh resolution. It was tested on the brain of seven healthy subjects at 7 T, with a final resolution of 1.5$\times$1.5$\times$5 mm\textsuperscript{3} for all maps acquired in 13 min. MNF could provide new fundamental insights into the inter-relationship between morphology (\textit{i.e.} tissue structure from the \hh maps) and physiology (\textit{i.e.} ion homeostasis from the \na maps) \textit{in vivo} to help improve our understanding of the human brain in general, and to study neuropathologies and their treatments. Since all \hhna MRF data is acquired simultaneously, all images are exactly co-registered with identical spatial and temporal resolutions. MNF could be useful in future longitudinal studies for assessing local time-dependent and conjoint \hhna MR changes during tasks or interventions. MNF was initially developed for neuroimaging, but it can be adapted to any other parts of the body.

\end{abstract}

\dates{Manuscript compiled on \today}

\begin{document}

\maketitle


\noindent 
Sodium ions (\naplusx) play an essential role in brain physiological processes such as cellular homeostasis (the maintenance of stable internal cell conditions of pH, temperature, ion concentration, and volume) and propagation of action potentials in neurons. The steady-state maintenance of the electrochemical gradient between intracellular and extracellular spaces is particularly important for ionic homeostasis. Well-defined transmembrane ionic gradients are necessary to absorb vital substrates (\textit{e.g.} glucose), to release products (\textit{e.g.} neurotransmitters), regulate intracellular metabolite concentrations (\textit{e.g.} ions), allow energy production (\textit{e.g.} oxidative phosphorylation), eliminate toxic byproducts (\textit{e.g.} amyloid oligomers), and control cell volume by regulation of the water osmotic pressure \cite{erecinska1994ions, lodish2000molecular}. Sodium homeostasis itself is regulated by the \nakx-ATPase (sodium-potassium pump) which requires about 50\% of the energy produced within the cells for normal brain function \cite{ames2000cns}, and therefore relies on an efficient and well maintained cellular energetic metabolism.

High-resolution imaging of sodium in the living human brain could therefore provide a wealth of new information leading to a better understanding of brain physiology in both health and disease \cite{madelin2022x}. It is possible to detect sodium ions \textit{in vivo} non-invasively with sodium magnetic resonance imaging (\na MRI), but so far low signal-to-noise ratio (SNR), long acquisition times and low resolution have been major hurdles for its application in clinical work or in basic brain research. Unlike conventional hydrogen (\hhx, proton) MRI, a cornerstone of modern-day healthcare and neuroscience, it is much more difficult to capture clear sodium images since the \na signal is about 20,000 times lower than the \hh signal in brain \cite{madelin2014sodium, gast2023recent}. Moreover, while \hh nuclei have a spin \(\frac{1}{2}\), leading to relatively simple dynamics producing an MR signal that persists tens or hundreds of milliseconds, \na nuclei have a spin \(\frac{3}{2}\) and undergo complicated quadrupolar dynamics resulting in an MR signal that disappears quickly within a few milliseconds, making it difficult to detect with standard MR techniques \cite{song202323na, madelin2013biomedical}. 

\begin{figure*}[t!]
    \centering
    \includegraphics[width=1\textwidth]{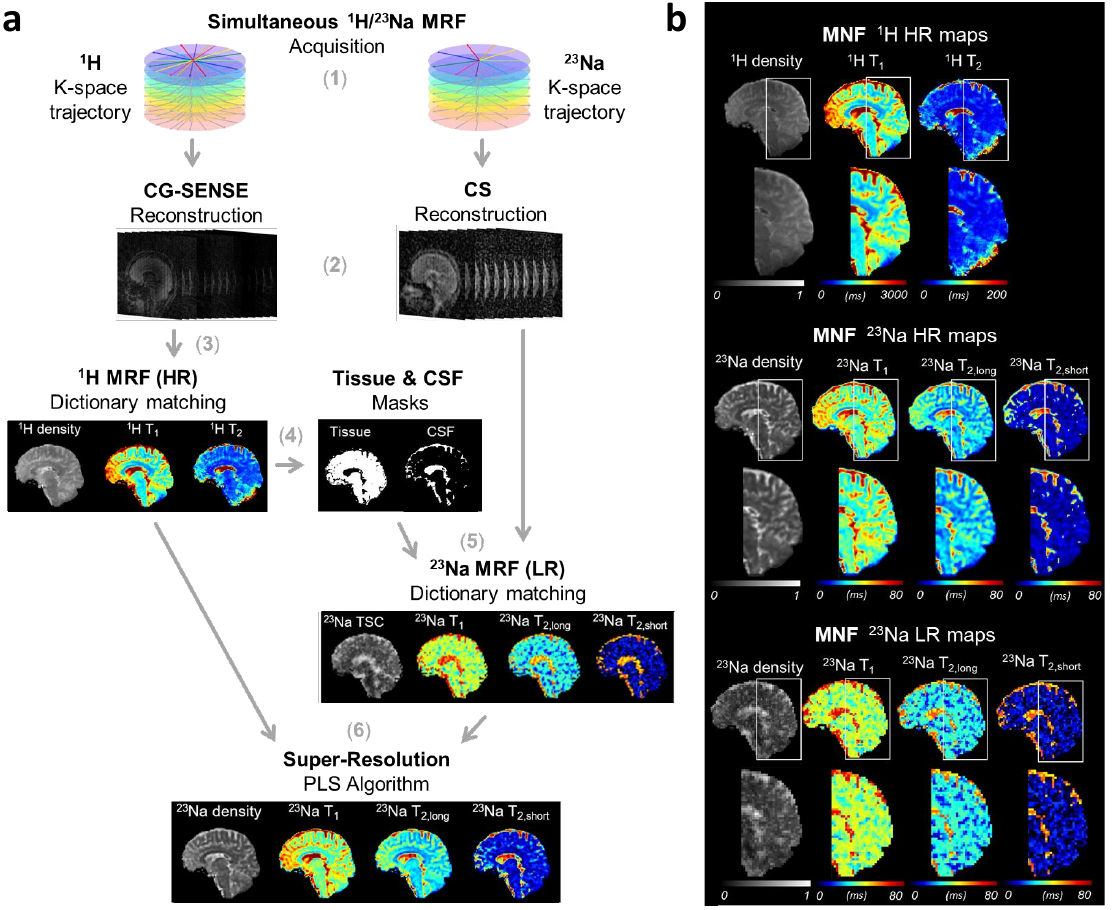}
    \caption{\textbf{The multinuclear fingerprinting (MNF) method.} (\textbf{a}) MNF consists of: (\textbf{1}) simultaneous \hh/\na MRF data acquisition; (\textbf{2}) separate reconstructions of HR \hh images (using CG-SENSE) and LR \na images (using CS) from data acquired after each pulse of their respective MRF acquisitions; (\textbf{3}) MR fingerprints dictionary matching of the \hh MRF images to generate quantitative HR maps of \hh density, and relaxation times T\textsubscript{1} and T\textsubscript{2}; (\textbf{4}) Brain tissue (gray + white matters) and CSF masking from the HR \hh MRF maps; (\textbf{5}) MR fingerprints dictionary matching of the \na MRF images to generate quantitative LR maps of \na density, and relaxation times T\textsubscript{1}, T\textsubscript{2,short} and T\textsubscript{2,long}, using the downsampled (LR) masks, and with different MR fingerprints dictionaries for fluids (CSF) and brain tissue; (\textbf{6}) Application of a PLS super-resolution algorithm optimized for these data to generate HR \na maps \cite{rodriguez2023super}. (\textbf{b}) Examples of sagittal slices from MNF from the quantitative 3D maps of \hh density, T\textsubscript{1} and T\textsubscript{2}, and \na density, T\textsubscript{1}, T\textsubscript{2,short} and T\textsubscript{2,long} in a healthy brain. The part of each map within the white rectangle is zoomed in below its respective map. For the \na data, we show the maps before (LR) and after (HR) application of the super-resolution algorithm. Resolution = 1.5$\times$1.5 mm\textsuperscript{2} for the HR slices, and 2.85$\times$2.85 mm\textsuperscript{2} for the LR slices. \textit{Abbreviations:} CG-SENSE = Conjugate Gradient Sensitivity Encoding; CS = Compressed Sensing; CSF = Cerebrospinal Fluid; PLS = Partial Least Square; LR = Low Resolution (2.85$\times$2.85$\times$5 mm\textsuperscript{3}); HR = High Resolution (1.5$\times$1.5$\times$5 mm\textsuperscript{3})}
    \label{fig_mnf_diagram_results}
\end{figure*}

In order to improve the temporal and spatial efficiency of \na MRI \textit{in vivo}, as well as its clinical significance in conjunction with \hh MRI, we developed a new imaging method called multinuclear fingerprinting (MNF) that  can provide quantitative images with both morphological (\hh from water) and physiological (\na from \naplusx) information in a single MR data acquisition. MNF consists of a new magnetic resonance fingerprinting (MRF) \cite{ma2013magnetic, cloos2016multiparametric} pulse sequence that can acquire and characterize both \hh and \na spin dynamics simultaneously to infer their respective local MR properties (density and relaxation times), combined with a specially-designed super-resolution (SR) algorithm. This MRF+SR process results in a set of \hh and \na maps of the whole brain with exact co-registration, as well as identical spatial and temporal resolutions.

\begin{figure*}[t]
    \centering
    \includegraphics[width=1\textwidth]{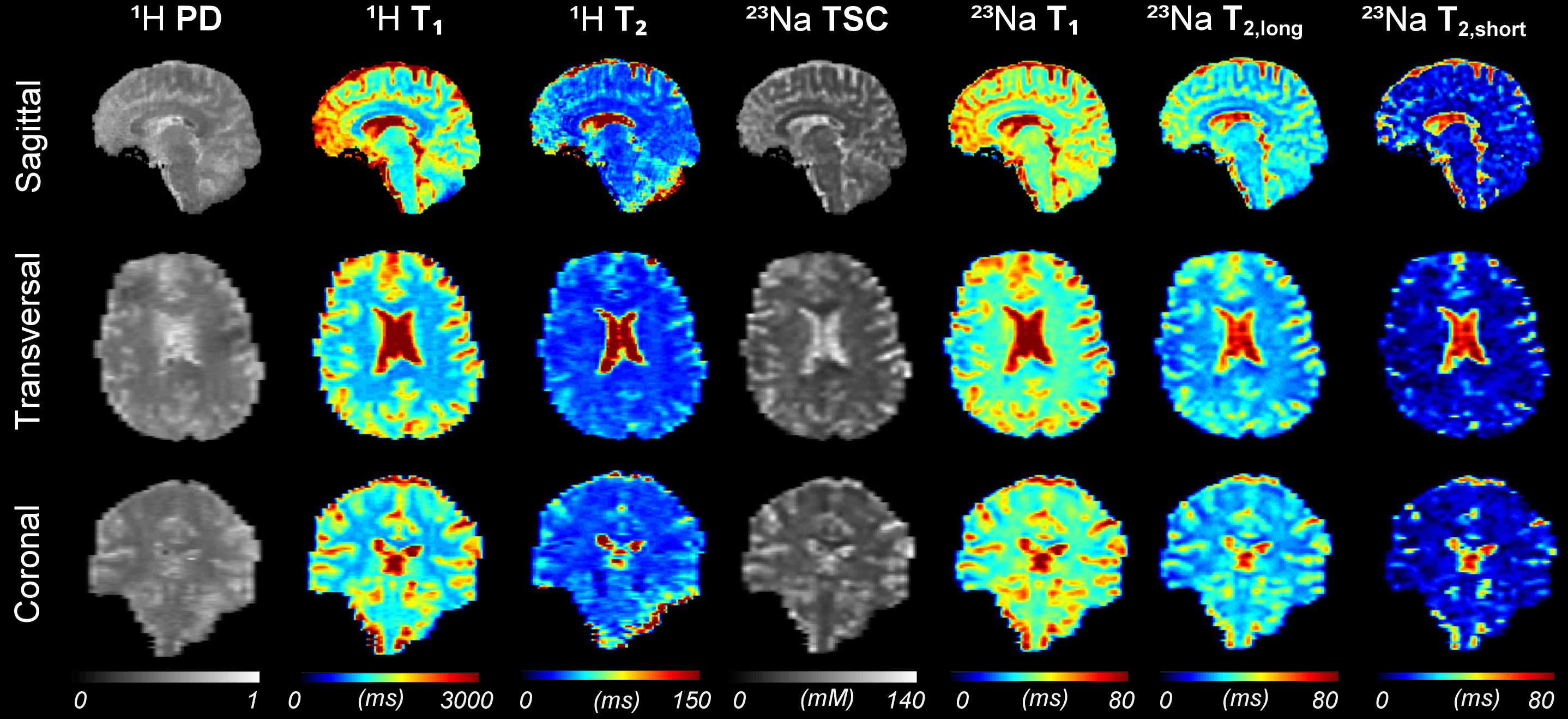}
    \caption{\textbf{Examples of the seven MNF HR \hh and \na maps of a whole brain.} Representative sagittal, transversal and coronal slices include proton density (PD), T\textsubscript{1} and T\textsubscript{2} from \hh, and tissue sodium concentration (TSC), T\textsubscript{1}, T\textsubscript{2,short} and T\textsubscript{2,long} from \nax. Resolution = 1.5$\times$1.5$\times$5 mm\textsuperscript{3}, TA = 13 min.}
    \label{fig_mnf_results}
\end{figure*}

In summary, to develop MNF, we first modified an ultra-high field (7 T) MRI system such that it can record MR signals from multiple nuclei simultaneously \cite{wang2021radially, yu2020simultaneous}. We then developed a new simultaneous 3D \hhna MRF pulse sequence, based on our previous work on simultaneous \hh MRF/\na MRI \cite{yu2020simultaneous, yu2022simultaneous, rodriguez2022repeatability}, to generate four "low-resolution" (LR, 2.85$\times$2.85$\times$5 mm\textsuperscript{3}) \na maps (tissue sodium concentration [TSC], \tonex, \ttwoshort and \ttwolongx), and three "high-resolution" (HR, 1.5$\times$1.5$\times$5 mm\textsuperscript{3}) \hh maps (proton density [PD], \tone and \ttwox) of the whole brain in a single 13-min acquisition. Finally, we implemented an iterative SR algorithm based on partial least square regression to model the voxelwise relationship between all the HR \hh and LR \na maps to derive the final HR \na maps on a subject-by-subject basis \cite{rodriguez2023super, van2015image}. The MNF pipeline is schematically summarized in Figure \ref{fig_mnf_diagram_results}. 

Since all MNF maps are acquired co-registered and simultaneously, they undergo the same local \bzero inhomogeneities, head motion, or pulsations from blood and cerebrosinal fluid (CSF), that can influence local MR signals. This could prove useful for: (1) for trying to reduce scan time by acquiring all multinuclear data in one single sequence instead of two consecutive sequences; (2) for studying the inter-relationship of brain morphology and physiology conjointly; and (3) for future dynamic or longitudinal studies to assess the simultaneous changes in local water and \naplus concentrations, and their respective relaxation times, during tasks or interventions.

We expect that the possibility to provide new \nax-based physiological maps alongside morphological \hhx-based maps with MNF will enable scientists and clinicians to directly assess ion homeostasis linked to cellular energy processes and related morphological tissue changes, bridge the gap in resolution that has held back our ability to study metabolism in the human brain \textit{in vivo}, and help better understand brain pathologies or monitor therapies. Although it was initally developed for neuroimaging, MNF can be adapted and optimized for any other parts of the body.

\vfill

\begin{figure*}[th]
    \centering
    \includegraphics[width=1\linewidth]{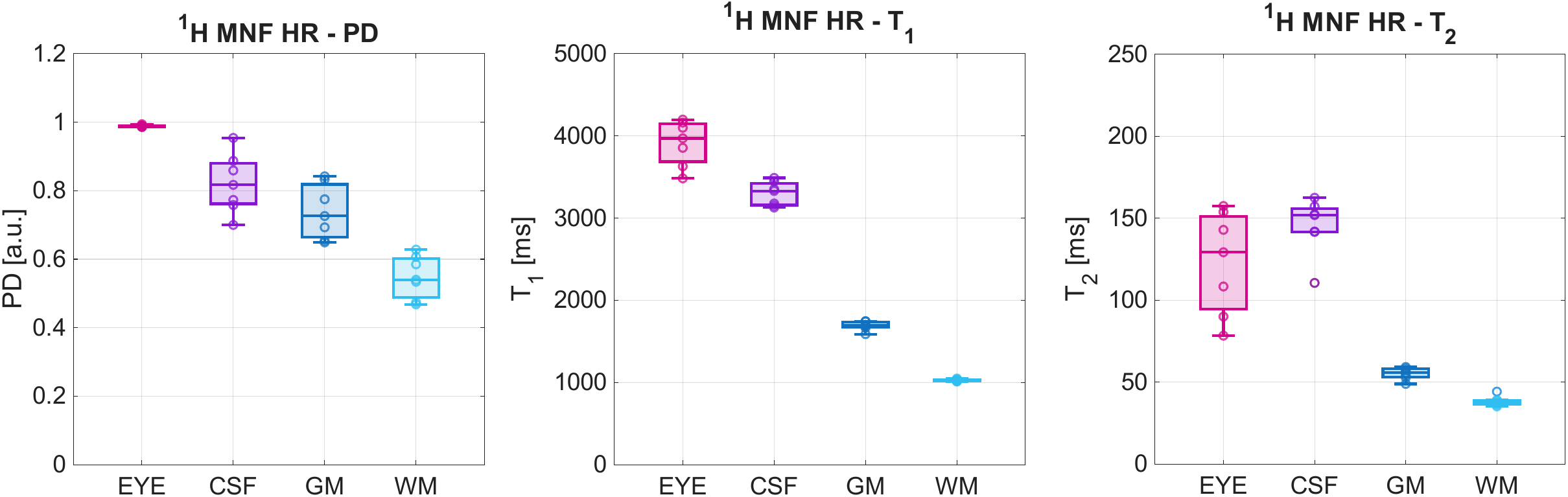}
    \caption{\textbf{\hh MNF HR measurements.} The boxplots show distributions and median values of final mean values (from the standard method) of the mean measurements from the N\textsubscript{ROI} and N\textsubscript{REF} values (histograms) from N = 7 subjects, from the \hh MNF HR data (PD, T\textsubscript{1}, T\textsubscript{2}), in the EYE, CSF, GM and WM ROIs. Highest values used were N\textsubscript{ROI} = 5\% to 100\% by steps of 5\% (20 values), N\textsubscript{REF} = 5\% to 100\% by steps of 5\% (20 values).}
    \label{fig_boxplot_1h_hr_20x20}
\end{figure*}

\section*{Results}

\subsection*{MNF maps}

Figure \ref{fig_mnf_results} shows representative examples of the seven 3D MNF HR \hh and \na maps in a healthy brain. These maps were all acquired simultaneously in 13 min and all have a spatial resolution of 1.5$\times$1.5$\times$5 mm\textsuperscript{3} after application of the SR algorithm. Notice that the slight radial undersampling artifacts noticeable in the frontal lobe area on the PD and \hh \ttwo sagittal maps did not affect the \na HR maps generated by the SR algorithm. Pre-processed full datasets from \hh MRF (PD, \tonex, \ttwo and \boneplusx) are shown in Figure S1, and full datasets from \na MRF (sodium density [SD], \tonex, \ttwoshortx, \ttwolongx, \deltaboneplus and \deltafzerox) are shown in Figure S2 in Supplementary Information. Only the seven maps of PD, \tonex, \ttwo (from \hh MRF) and TSC (derived from SD), \tonex, \ttwoshort and \ttwolong (from \na MRF) are shown as meaningful results with this method. The \boneplusx (from \hh MRF), \deltaboneplus and \deltafzero (from \na MRF) parameters were only included in the simulation models used to calculate the respective fingerprint dictionaries as complementary information used to improve the estimation of the seven final MNF maps.   

\subsection*{Measurements}

In the following sections, the \textit{final} mean and standard deviation (std) correspond to the mean $\pm$ one std value of the measurements calculated for each subject using three different statistical methods (standard, jackknife, booststrap) as described in details in "Methods/Statistical analysis", while the \textit{overall} mean corresponds to the mean $\pm$ one std of the final mean and of the final std values calculated over all subjects for each statistical method.

Figure \ref{fig_boxplot_1h_hr_20x20} shows boxplots of the final mean values of the \hh MNF HR data from N = 7 subjects, calculated using the standard statistical method. This data includes PD, \tonex, and \ttwo measured in four regions-of-interest (ROI): the vitreous humor of the eyes (EYE), cerebrospinal fluid (CSF), gray matter (GM) and white matter (WM). We can see that PD measurements present quite a large variability between subjects, except for PD in the EYE which was used as a reference. Relaxation times values in EYE and CSF are also quite variable, while \tone and \ttwo in GM and WM were pretty consistent with each other for all subjects.

\begin{figure*}[th]
    \centering
    \includegraphics[width=1\linewidth]{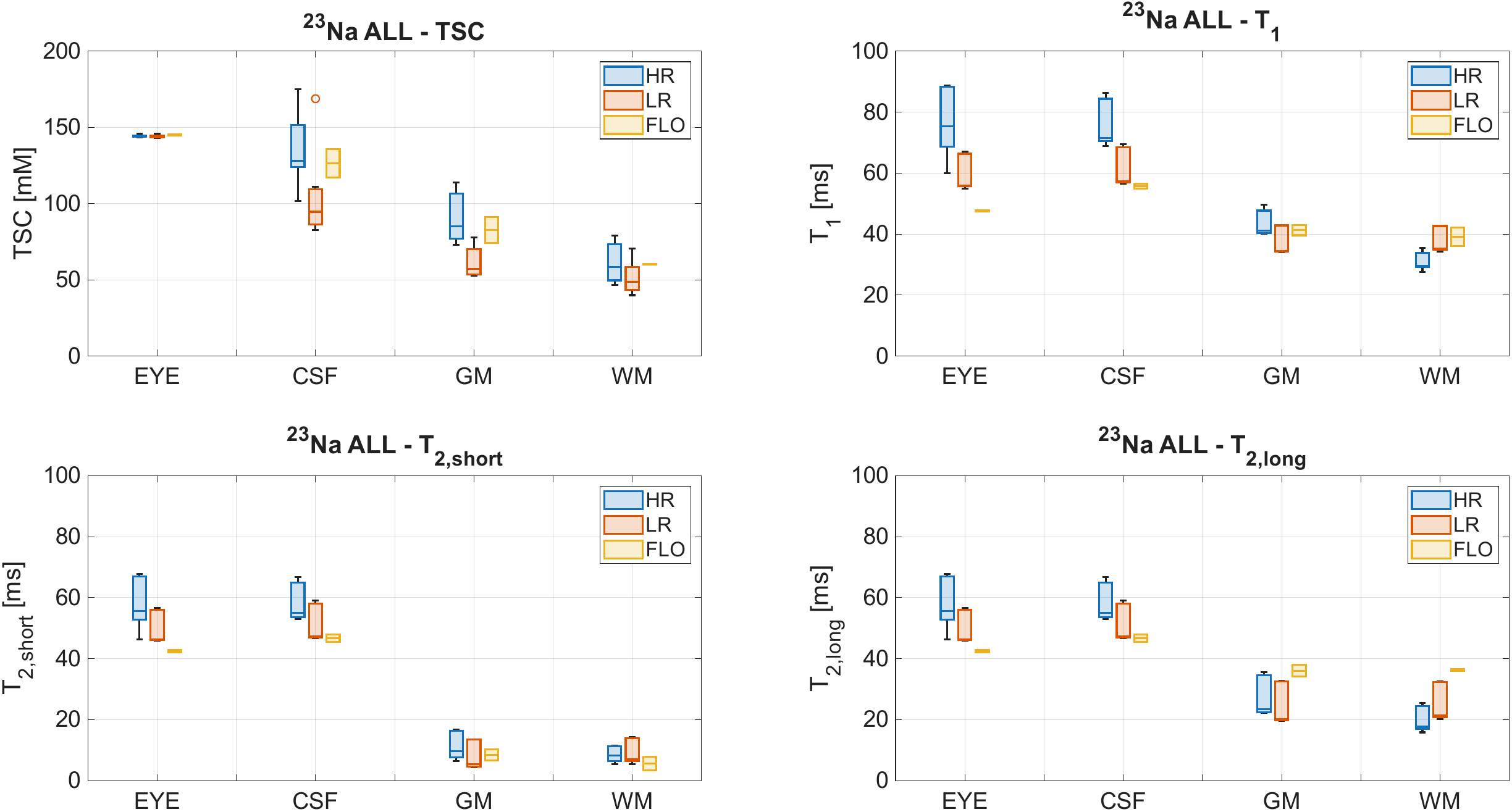}
    \caption{\textbf{\na MNF HR, LR, and FLORET measurements.} The boxplots show distributions and median values of the final mean values (from the standard method) of the mean measurements from the N\textsubscript{ROI} and N\textsubscript{REF} values (histograms), from the \na MNF HR (7 subjects), LR (7 subjects) and FLORET (FLO, 2 subjects: 5 and 6) data (TSC, T\textsubscript{1}, T\textsubscript{2,short}, T\textsubscript{2,long}), in the EYE, CSF, GM and WM ROIs. Highest values used were N\textsubscript{ROI} = 5\% to 100\% by steps of 5\% (20 values), N\textsubscript{REF} = 5\% to 100\% by steps of 5\% (20 values).}
    \label{fig_boxplot_23na_hr_lr_flo_all_20x20}
\end{figure*}

Figure \ref{fig_boxplot_23na_hr_lr_flo_all_20x20} shows boxplots of the final mean values of the \na MNF HR and LR data from N = 7 subjects, and of the FLORET (FLO) data acquired as a reference method on two subjects, all calculated using the standard statistical method. This data includes TSC, \tonex, \ttwoshortx, and \ttwolong measured in the EYE, CSF, GM and WM. Overall, \na MNF HR seems to overestimate TSC and relaxation time values compared to \na MNF LR and FLORET data, except for WM relaxation times.

Table \ref{tab_summary_mean_values_mnf_brain} (placed after References) presents a summary of the overall means and standard deviations from \hh MNF HR, \na MNF HR and LR in the EYE, CSF, GM and WM. See Tables S1-S3 in Supplementary Information for a detailed summary of all the mean value measurements in each subject. We can see that the overall mean values are identical for all three statistical methods, but that the overall standard deviations (\textit{i.e.} the mean uncertainty of the measurements overt all subjects) are different: of the order of 12.49$\pm$7.85\% (range [3.22\%, 32.92\%]) for the standard method, of the order of 0.03$\pm$0.03\% (range [0.00\%, 0.12\%]) for jackknife, and of the order of 0.63$\pm$0.40\% (range [0.16\%, 1.66\%]) for bootstrap. 

Individual measurements for each subject are presented in Supplementary Information. Figures S3-S25 show measurement histograms for single values of N\textsubscript{REF} (80\%) and N\textsubscript{ROI} (90\%). Figures S26-S31 show boxplots of the mean values measured in all seven subjects with single values of N\textsubscript{ROI} (80\%) and N\textsubscript{REF} (90\%). Figures S32-S54 show histograms of mean values from multiple measurements with both N\textsubscript{REF} and N\textsubscript{ROI} ranging from 5\% to 100\% by steps of 5\%. Figures S55-S59 show boxplots of the mean values of the mean measurements using N\textsubscript{ROI} and N\textsubscript{REF} from 5\% to 100\% by steps of 5\%. 

\subsection*{Validation}

Table \ref{tab_relaxation_1h_comparison} (placed after References) presents a comparison of the overall mean values (standard statistical method) from \hh MNF HR measured \textit{in vivo} in brain at 7 T with values from the literature, in EYE, CSF, GM and WM. We can see that mean PD and \hh relaxation times measured with \hh MNF HR are within or close to (within one std) the ranges of values from the literature, which is particularly sparse for the EYE and CSF regions. The two main discrepancies are: (1) PD in CSF, which is quite low (0.81 $\pm$ 0.09) compared to an expected value of 0.99; and (2) \ttwo in CSF, which is higher than the value from a unique reference. 

Table \ref{tab_relaxation_23na_comparison} (placed after References) presents a comparison of the overall mean values (standard statistical method) from \na MNF LR and HR measured \textit{in vivo} in brain at 7 T with values from the literature, in EYE, CSF, GM and WM. Mean values measured from \na MNF LR are within the range of values from the literature, except \tone and \ttwoshort in GM, and \ttwoshort in WM, which are slightly above the range but still within one standard deviation. Most mean values measured from \na MNF HR are in the higher part of, or above, the range of values from the literature, except WM \tone which is slightly below this range.

The Wilcoxon rank-sum test did not detect any statistically significant difference (after Bonferroni correction) between the \na MNF LR, \na MNF HR and \na FLORET measurements in GM, WM CSF and EYE, except in four cases, all of them only between HR and LR data: \tone in EYE, CSF and WM, and TSC in GM (see Table S5 in Supplementary Information).

\section*{Discussion}

Overall, both the \hh MNF HR and the \na MNF LR data reconstructed directly from the simultaneous \hhna MRF acquisition provided mean values of PD, TSC and relaxation times from both nuclei that were similar to the values found in the sparse literature in healthy brain at 7 T. However, the application of the SR algorithm to the \na MNF LR data to generate the HR maps had a tendency to slightly increase the values compared to literature. This could be explained by the reduction of partial volume effects (PVE) in the HR data combined with the fact that all the HR \na values are compared with values from the standard \na MRI literature, which relies on data acquired with a low resolution similar to our \na LR data (or even lower). 

While absolute values measured in each healthy volunteer can be of interest for the diagnosis of brain pathologies, the main goal of MNF would be its application into longitudinal studies where the subjects will be scanned multiple times either during tasks, therapies, or simply to assess the potential evolution of pathologies over time without intervention. Quantitative changes in all these MNF metrics should provide more relevant information about brain morphological and physiological alterations in different conditions than the initial absolute value itself.

The systematic use of multiple values of N\textsubscript{REF} highest voxel values for quantifying PD and TSC and of N\textsubscript{ROI} for measuring mean values of all metrics in different ROIs can reduce the sensitiviy of these measurements to subjective operator-dependent manual drawing of ROIs and to PVE at the borders of these ROIs. It should thus provide a more robust and automated quantification of MNF metrics in different regions of the brain. In addition, the application of resampling methods such as jackknife or bootstrapping can also significantly reduce uncertainties on these ROI-based quantitative estimates for each subject, where the mean std of the mean value measurements can go from more than 12\% using standard statistics to less than 1\% with resampling.     

The simultaneous acquisition of \hh and \na MRF data was made possible on our scanner at 7 T thanks to the insertion of an external frequency generator (local oscillator) in the radiofrequency (RF) cabinet of the system. However, such hardware modification can be difficult to carry out on clinical scanners at 3 T or 7 T. In that case, an alternative version of MNF needs to be developed with interleaved \hhna MRF data acquisition instead of truely simultaneous \cite{meyerspeer2016simultaneous, lopez2022interleaved}, where the signal of \hh and \na can be recorded through two different analog-to-digital converters (ADC) set up in close succession. In that case, \hh and \na MR raw data acquisitions will be delayed only by a few milliseconds and can be considered quasi-simultaneous. 

Further improvements of MNF to increase spatial resolution in all directions and reduce total scan time will include stack-of-spirals or 3D spiral (FLORET-type) k-space trajectories to increase SNR efficiency, combined with regularization-by-denoising (RED) implementation in the iterative image reconstruction process \cite{romano2017little, qiusheng2020compressed} for increasing SNR furthermore, as well as deblurring the \na data.

In conclusion, we developed the MNF method to generate seven whole-brain maps related to morphology (\textit{i.e.} tissue structure from three \hh MRF maps) and physiology (\textit{i.e.} sodium homeostasis from four \na MRF maps), all acquired simultaneously with exact co-registration, and identical spatial and temporal resolutions. Thus far, the MNF data can be acquired over the whole brain in 13 min at 7 T with a final spatial resolution of 1.5$\times$1.5$\times$5 mm\textsuperscript{3}. We expect that MNF could prove useful for two purposes: (1) as a clinical tool to monitor the evolution of brain pathologies and their responses to therapy, or (2) as a basic research imaging tool to study conjointly brain morphology and physiology \textit{in vivo}, and their interaction in healthy brain or in pathologies. This proof-of-concept for MNF was first implemented in brain, but it can also be adapted to other parts of the body.

\begin{figure*}[th]
    \centering
    \includegraphics[width=1\textwidth]{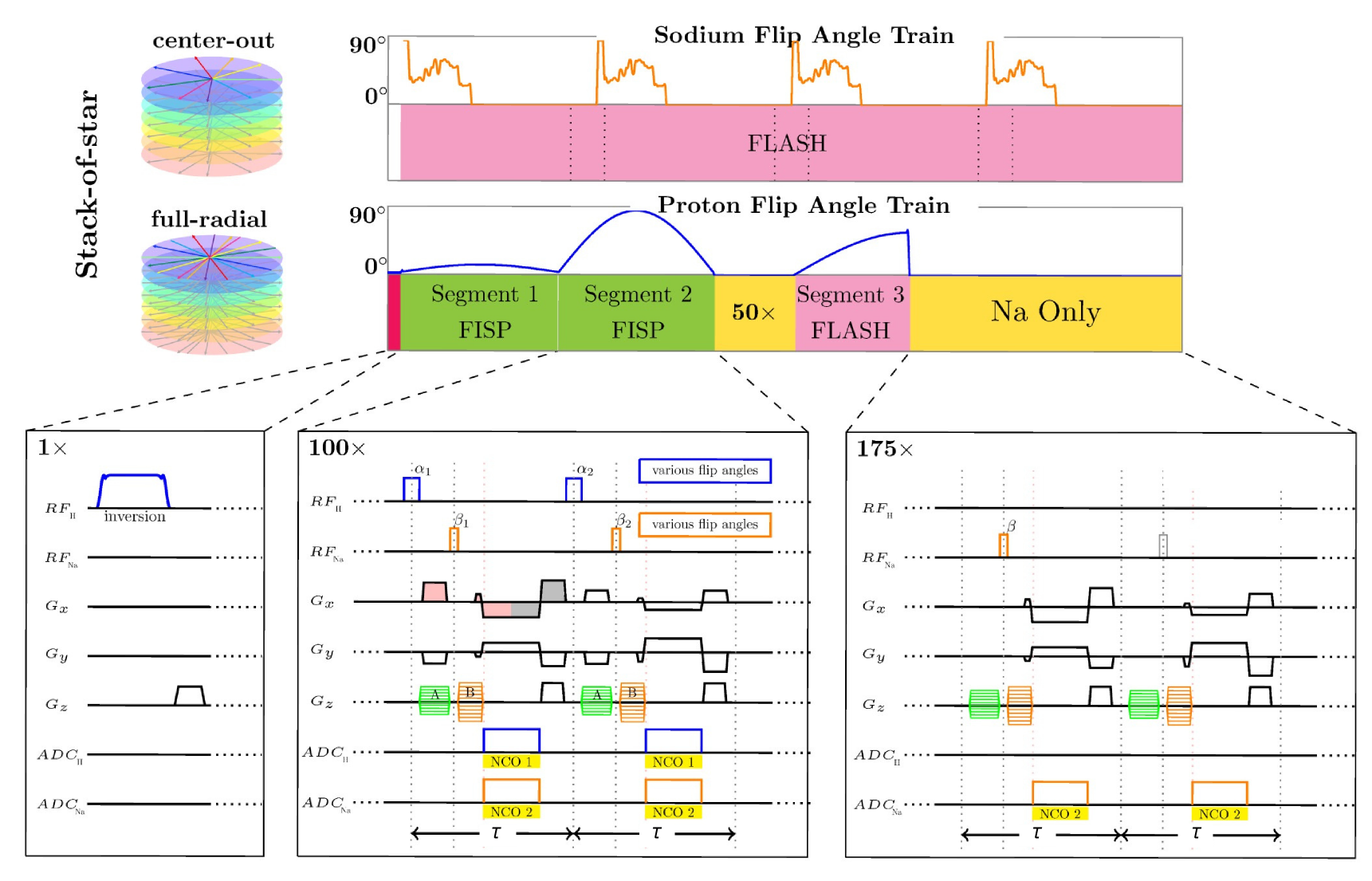}
    \caption{\textbf{The simultaneous \hhna MRF pulse sequence.} Magnetic resonance fingerprinting (MRF) data is acquired simultaneously for proton (\hhx) using a full-radial stack-of-stars k-space trajectory and for sodium (\nax) using a center-out stack-of-stars trajectory. The \hh flip angle (FA) train is made up of 500 RF pulses $\alpha$ applied every delay $\tau$ of 7.5 ms, and that includes 27 variable pulses (for data acquisition) and 225 pulses with FA = 0$^\circ$ (no data acquisition). The \na FA train is made up of 22 variable RF pulses $\beta$ applied every 2$\tau$ (with following pulses set to zero), applied 4 times per shot. Slice encoding gradients A (green) and B (orange) are used to encode the same slice for different nuclei. TR per shot was 3.75 s. \textit{Abbreviations:} FISP = Fast Imaging with Steady-state Free Precession; FLASH = Fast Low Angle Shot; NCO N = Numerical Oscillator event N.} 
    \label{fig_mnf_seq}
\end{figure*}

\section*{Methods}

\subsection*{Hardware} 

The experiments were performed at 7 T (MAGNETOM, Siemens, Erlangen, Germany) using a 16-channel transmit/receive (8 \hh channels and 8 \na channels) dual-tuned \hhna RF head coil developed in-house \cite{wang2021radially}. The scanner was modified to allow simultaneous acquisition of both \hh and \na signals by inserting an external frequency generator in the RF cabinet of the system to demodulate the \na signal with a proper local oscillator \cite{meyerspeer2016simultaneous, yu2020simultaneous}.

\subsection*{Brain imaging protocol}

MNF brain images were acquired on seven healthy volunteers (4 females, mean age = 27±3 years) after informed consent, in accordance with the New York University Grossman School of Medicine institutional review board and national guidelines. All volunteers were scanned with the simultaneous 3D \hhna MRF acquisition, and their \na MNF LR maps were then post-processed with the super-resolution algorithm to generate \na MNF HR maps that match the resolution of the \hh MNF HR maps. In addition, two subjects (\#5 and \#6) participated in a second session where standard measurements of \tone using multi-TR \na FLORET saturation-recovery (SR) experiments, and of \ttwoshort and \ttwolong using multi-TE \na FLORET experiments, were acquired for comparison with the MNF results. 

\subsection*{MNF data acquisition} 

\subsubsection*{Simultaneous 3D \hhna MRF}

We developed the new simultaneous 3D \hhna MRF pulse sequence based on our previous work on simultaneous \hh MRF/\na MRI with a stack-of-star k-space trajectory \cite{yu2020simultaneous, yu2022simultaneous, rodriguez2022repeatability}. It was optimized to generate low resolution (LR, 2.85$\times$2.85$\times$5 mm\textsuperscript{3}) \na maps, and high resolution (HR, 1.5$\times$1.5$\times$5 mm\textsuperscript{3}) \hh maps in the whole brain at 7 T in about 13 min. In this pulse sequence, the nuclear spins are sequentially excited every inter-pulse delay $\tau$ for \hh, and every inter-pulse delay $2\tau$ for \na (delay between the start of each pulse), using non-selective RF pulses followed by one simultaneous ADC readout for both nuclei. The partition phase-encoding gradient moments were distributed such that images from both nuclei had the same slice thickness. The frequency-encoding gradient moments were distributed such that a full radial trajectory for \hh and a center-out radial trajectory for \na were obtained in k-space, leading to a ratio of approximately 1.9 for in-plane resolution between the \hh and \na images, which corresponds to half of the ratio of their respective gyromagnetic ratios \cite{yu2022simultaneous}. Fig. \ref{fig_mnf_seq} shows a schematic diagram of the simultaneous 3D \hhna MRF pulse sequence.

\subsubsection*{\hh RF pulse train}

The \hh 500-pulse train is the same as the one introduced by Yu. et al \cite{yu2022simultaneous}. It starts with an adiabatic inversion pulse \cite{ordidge1996frequency}, followed by 275 rectangular RF pulses with variable flip angle (FA) and constant pulse duration of 1 ms, distributed in different segments, and some extra delays between the segments (225 pulses with FA = 0$^\circ$). All pulses are separated by a delay $\tau$. This pulse train contains both fast imaging with steady precession (FISP) and fast low angle shot (FLASH) segments to help differentiate \tonex, \ttwox, and \boneplus effects \cite{cloos2016multiparametric}. The first segment (FISP, pulses 1–100) includes flip angles up to 20$^\circ$. The second segment (FISP, pulses 101–200) contains higher flip angles (up to 90$^\circ$) to help encode \ttwox. A 50-$\tau$ delay corresponding to pulses 201-250 with FA = 0$^\circ$ was inserted to recover some magnetization before the third segment with flip angles up to 60$^\circ$ (FLASH, pulses 251–325), which was RF-spoiled to help encode \boneplusx. A final 175-$\tau$ delay corresponding to pulses 326–500 with with FA = 0$^\circ$ was added at the end of the pulse train to allow \tone recovery for the \hh magnetization before the next inversion pulse and pulse train acquisition, resulting in a repetition time (TR) of 3.75 s per "shot" (one set of 275 data acquisitions from the 275-pulse variable train and including the 225-pulse train with FA = 0$^\circ$).

\subsubsection*{\na RF pulse train}

The \na pulse train consists of 22 rectangular RF pulses of duration 0.6 ms and separated by a delay of $2\tau$. The 22 flip angles were optimized by a genetic algorithm to minimize the Pearson correlation coefficient between signals of GM and WM matter based on the simulated signal evolution for \na nuclear spins, and using average \na relaxation times for GM and WM from the literature, similarly to the method described in O'Donnell et al. \cite{odonnell2023mapping, o2025correlation}. Note that in O'Donnell et al. \cite{odonnell2023mapping, o2025correlation}, a 23-pulse train was optimized for \na MRF-only acquisition, and which included a composite 90$^\circ$-180$^\circ$-90$^\circ$ inversion block at the beginning of the pulse train. After multiple experiments with the 23-pulse train, and in order to significantly reduce specific absorption rate (SAR) from our simultaneous \hhna MRF sequence, we decided to remove the 180$^\circ$ RF pulse from the composite inversion block, and resimulate the \na MRF dictionary with the resulting 22-pulse train. Modelization of the \na nuclear spin $\frac{3}{2}$ dynamics under this 22-pulse train was performed using the irreducible spherical tensor operator (ISTO) formalism \cite{madelin2014sodium} in Matlab (The MathWorks Inc, Natick, MA, USA). The resulting 22 \na FA excitations were: 90$^\circ$, 90$^\circ$, 39$^\circ$, 32$^\circ$, 35$^\circ$, 38$^\circ$, 35$^\circ$, 49$^\circ$, 42$^\circ$, 52$^\circ$, 62$^\circ$, 45$^\circ$, 61$^\circ$, 63$^\circ$, 60$^\circ$, 55$^\circ$, 52$^\circ$, 53$^\circ$, 30$^\circ$, 27$^\circ$, 27$^\circ$, 28$^\circ$. All RF pulse phases were equal to zero. A final delay of 600 ms was then applied at the end of the pulse train to allow full \tone recovery for the \na magnetization before the next pulse train acquisition, and such that this \na pulse train can be applied 4 times per shot (TR = 3.75 s).

\subsubsection*{Acquisition parameters}

The simultaneous 3D \hhna MRF data was acquired in brain with the following parameters: Field-of-view (FOV) = 240$\times$240$\times$175 mm\textsuperscript{3}, in-plane sagittal resolution =  1.5$\times$1.5 mm\textsuperscript{2} for \hh and 2.85$\times$2.85 mm\textsuperscript{2} for \nax, inter-pulse delay $\tau$ = 7.5 ms for \hh and $2\tau$ = 15 ms for \nax, echo time (TE) = 2 ms for \hh and 1.2 ms for \nax, 1 slab of 35 slices, sagittal slice thickness = 5 mm for both \hh and \nax, 3 shots per slab, TR = 3.75 s/shot, 2 acquisitions averaged together, total scan time = 13 min.

\subsection*{MNF data reconstruction} 

\subsubsection*{\hh MRF}

\paragraph{Image reconstruction:} 

For the \hh MRF data, all individual images acquired after each RF pulse were reconstructed using conjugate gradient sensitivity encoding (CG-SENSE) \cite{pruessmann2001advances} to reduce the radial artifacts. The radial samples from the 275 data acquisitions were reconstructed with an 11-acquisition sliding window \cite{cao2017robust}. In total, 265 image frames were therefore reconstructed by sliding this 11-acquisition window in increments of 1 acquisition. The fingerprint dictionary was grouped and averaged with the same sliding window as the CG-SENSE reconstruction along the time domain \cite{cloos2016multiparametric}. 

\paragraph{Fingerprint dictionary:} 

The \hh fingerprint dictionary was computed using the extended phase graph formalism \cite{weigel2015extended, cloos2016multiparametric} implemented in C++, with different step sizes: \tone ranged from 150 to 4347 ms, and \ttwo ranged from 15 to 435 ms, both incremented in steps of 5\%; \boneplus ranged from 10$^\circ$ to 130$^\circ$, in steps of 1$^\circ$ \cite{yu2022simultaneous}.

\paragraph{Matching:} 

Matching between the acquired signal and the fingerprint dictionary was performed voxel by voxel. The final \tonex, \ttwo and \boneplus maps were obtained from the maximum correlation between the dictionary and pixel signal evolution, where the correlation was defined as the matrix product between them \cite{yu2022simultaneous}. 

\subsubsection*{\na MRF}

\paragraph{Image reconstruction:} 

For \na MRF, all individual images acquired after each RF pulse were reconstructed using compressed-sensing (CS) using total-generalized-variation regularization to maximize SNR and minimize radial artifacts \cite{knoll2011second}. As the radial spokes were acquired with golden angle increment, the radial samples from the 4 pulse trains from each shot were reconstructed altogether. The radial samples from 22 acquisitions per pulse train were reconstructed with a 5-acquisition circular sliding window to increase SNR. In total, 22 image frames were then reconstructed by sliding this window in increments of 1 acquisition. The fingerprint dictionary was grouped and averaged with the same sliding window as the CS reconstruction \cite{cloos2016multiparametric}. 

\paragraph{Fingerprint dictionary:} 

The spin $\frac{3}{2}$ dynamics simulation to generate the \na fingerprint dictionary was performed in Matlab R2020b. Signals were simulated starting from thermal equilibrium and propagated under the optimized 22-pulse  train using the irreducible spherical tensor operators (ISTO) framework \cite{madelin2014sodium, kratzer20213d, o2025correlation}. Parameter ranges to build the dictionary were \tone from 20 ms to 74 ms (steps of 2 ms), \ttwolong from 10 ms to 66 ms (steps of 2 ms), \ttwoshort from 0.5 to 66 ms (steps of 0.5 ms for \ttwoshort from 0.5 to 2 ms, then steps of 2 ms), \deltaboneplus factor from 0.7 to 1.3 (applied as a multiplying factor to the RF amplitude, steps of 0.1), and \deltafzero from -60 Hz to 60 Hz (steps of 10 Hz). Parameter combinations where \ttwolong $>$ \tone and \ttwoshort $>$ \ttwolong were omitted from the computation.

\paragraph{Matching:}  Matching between the acquired signal and the fingerprint dictionary was performed voxel by voxel. The brain was first segmented into two compartments (WM+GM and EYE+CSF) using the masks from the segmentation processing described in section "Brain segmentation" below. For EYE+CSF, the following matching constraints were implemented: \tone in range [49-74] ms, \ttwolong in range [42-66] ms and \ttwoshort = \ttwolong $\equiv$ \ttwox. For GM+WM, \tone was in range [20-48] ms, \ttwolong in range [10-40] ms and \ttwoshort in range [0.5-20] ms. These ranges were chosen in accordance with the expected values in these tissues from the literature \cite{o2025correlation}. The fingerprint dictionary (size = 113,880 entries for GM+WM and 5,187 for EYE+CSF) was matched voxelwise to the \na signals using Pearson correlation. 

\paragraph{Correlation weighting:} 

Because of the dictionary size and due to the low intrinsic SNR of the \na images, and thus noisy \na signal evolution over the \na RF pulse train, it was therefore possible that more than fingerprint could generate a high correlation for a single voxel signal evolution. To account for this and to increase robustness of the matching process, we included matches for a subset of the top correlation coefficients for each voxel $v$, and constructed the final \na LR maps by calculating the correlation-weighted parameter $X_{v}$ from the dictionary of values $X_{v,i}$, and corresponding to the matching correlation coefficient $W_{v,i}$ according to \cite{o2025correlation}: 
\begin{equation}
    X_{v} = \frac{\sum^{k}_{i=1} W_{v,i}X_{v,i}}{\sum^{k}_{i=1} W_{v,i}},
    \label{eq:correlation_weighting} 
\end{equation}
where $k$ was the maximum number of correlation coefficients used for weighting, and $X_{v}$ and $X_{v,i}$ correspond to the metrics \tonex, \ttwoshortx, \ttwolongx, \deltaboneplus factor or \deltafzerox. 

The number $k$ of highest correlation coefficients used for the weighting was determined empirically considering the number of dictionary entries and the SNR of the \na data, after testing a wide range of values for $k$ ($k$ = 1 to 1,000), and chosen to be $k$ = 50 for EYE+CSF ($\sim$1\% of the total number of correlation coefficients calculated from all dictionary entries for these tissues) and $k$ = 500 for GM+WM ($\sim$0.5\% of the total number of correlation coefficients calculated from all dictionary entries for these tissues).

\subsubsection*{Brain segmentation} 

The MNF brain data from each volunteer was segmented into four ROIs using SMP12 (UCL, London, UK) in Matlab \cite{ashburner2014spm12}: vitreous humor from the eyes (EYE), grey matter (GM), white matter (WM), and cerebrospinal fluid (CSF). Tissue segmentation was performed using the PD, \tonex, and \ttwo maps from the \hh MNF HR data as input images. Note that SPM12 generate five probability maps by default: GM, WM, CSF, SKULL, and OTHER tissues excluding the eyes. The eye segmentation was generated by calculating EYE = 1 - [CSF + GM + WM + SKULL + OTHER] probabilities. The EYE, GM, WM and CSF probability maps were then normalized and binarized with a threshold of 0.90 to minimize the number of voxels with multiple tissue components to generate non-overlapping EYE, GM, WM, and CSF HR masks. LR masks were also calculated by resizing the HR masks in Matlab using the function \textit{imresize3} with \textit{nearest} option.

\subsubsection*{PD and TSC quantification}

\paragraph{Proton density (PD):} 

Initial non-normalized PD maps were calculated voxelwise as the mean value of the division between the signal evolution and the optimal fingerprint from dictionary matching. The bias field from these inital maps was corrected by bias regularization in SPM12 \cite{ashburner2014spm12, ashburner2013symmetric} during the segmentation process for generating different tissues masks. The resulting images were then normalized by the mean value measured in the vitreous humor of the eyes, accounting for a reference PD in the eyes of 0.98 \cite{chau2014osteogenesis, forrester2015eye} to generate truely quantitative final PD maps in the brain. In order to make the final PD measurements more accurate and more robust against potential local map inhomogeneities with the reference ROI, as well as PVE from the edges of the reference ROI, PD quantification was performed 20 times by using the mean reference value from only the N\textsubscript{REF} \% highest voxel values of the reference ROI (EYE), with N\textsubscript{REF} = 5\% to 100\% by steps of 5\% (20 N\textsubscript{REF} values). The mean PD values over these 20 PD maps can then be measured, as explained in more details in subsection "Statistical analysis". 

\paragraph{Tissue sodium concentration (TSC):} 

Multiple sodium density (SD) maps were first calculated as the mean value of the division between the absolute value of the \na signal evolution for each voxel $v$ and the $k$ optimal fingerprints from dictionary matching. Correlation-weighting was then performed according to Equation \ref{eq:correlation_weighting}, where $X_{v,i}$ was the unweighted SD and $W_{v}$ was the weighted SD. The resulting correlation-weighted SD map was then corrected for \boneminus inhomogeneities (coil sensitivities) \cite{axel1987intensity}: The map was divided by the normalized magnitude \na image of a uniform phantom previously acquired with the same RF coil. The phantom was a PET screw cap bottle with diameter = 28 cm, height = 50 cm, and volume = 22.7 L (which filled most of the coil volume), and filled with an aqueous solution with 150 mM NaCl. The phantom uniform image was acquired with FLORET with the following parameters: 4 acquisitions averaged together, FOV = 320 mm isotropic, resolution = 10 mm isotropic, TE = 1 ms, TR = 100 ms, 32 averages/acquisition, 3 hubs at 45$^\circ$, 100 interleaves/hub, flip angle = 80$^\circ$, RF pulse duration = 2000 $\mu$s, total acquisition time = 16 min/acquisition. All 4 complex images were reconstructed as described in section "\na FLORET data reconstruction", and averaged together. The magnitude of the final averaged phantom image was then normalized by its maximum value, and its inverse image was used to correct the brain SD map. 

The \boneminusx-corrected SD map was then normalized by the mean value over the vitreous humor of the eyes from the EYE mask (accounting for a water content of 0.98) and then multiplied by a reference TSC value for the eyes of 145 mM \cite{kokavec2016biochemical, winkler1989regional, adlung2022quantification} to generate the TSC map. Similarly to PD, in order to make the final TSC measurements more accurate and more robust against potential local map inhomogeneities with the reference ROI, as well as PVE from the edges of the reference ROI, TSC quantification was performed 20 times by using the mean reference value from the N\textsubscript{REF} \% highest voxel values of the reference ROI (EYE), with N\textsubscript{REF} = 5\% to 100\% by steps of 5\% (20 N\textsubscript{REF} values). The mean TSC values over these 20 TSC maps can then be measured, as explained in more details in subsection "Statistical analysis". 

\subsubsection*{Super-resolution (SR) post-processing} 

Finally, a super-resolution algorithm based on partial least square (PLS) regression \cite{van2015image, rodriguez2023super} between the \hh HR maps and the \na LR maps acquired with simultaneous \hhna MRF was applied to generate \na HR maps with the same resolution as the \hh HR maps. The algorithm is described in details in Rodriguez et al. \cite{rodriguez2023super}, and was simply adapted here to operate with all seven maps as inputs (3 HR \hh maps + 4 LR \na maps), instead of four images (3 HR \hh maps + 1 LR \na image). In summary, the goal of the algorithm is to link all \na metrics from the LR maps to a combination of \hh-based metrics from the HR maps by modeling the distribution of co-registered measurements from both types of images using PLS regression. An iterative loop, including deconvolution/convolution with point spread functions and image resizing between LR and HR with a fast Fourier transform approach, were included in the algorithm to generate the final HR \na maps without losing features from the LR \na maps. The average processing time was 5.1 s per slice on a computer with 6-core CPU, 3.4 GHz, 32 GB RAM. The output for this SR postprocessing was four \na HR maps. 

\subsection*{\na FLORET data acquisition}

\subsubsection*{Acquisition parameters}

For comparison with MNF results, \na data was also acquired in two subjects (\#5 and \#6) in different scanning sessions, with the Fermat looped, orthogonally encoded trajectories (FLORET) sequence \cite{pipe2011new} with the following parameters: 3 hubs at 45$^\circ$ with 240 interleaves per hub, with diﬀerent TR and TE values for relaxation fitting (see below), 4 averages, 3.75 mm isotropic resolution, FOV = 240$\times$240$\times$240 mm\textsuperscript{3}, 64 slices, rectangular RF excitation pulses with FA = 90$^\circ$ and pulse duration = 0.8 ms. 

\subsubsection*{Multi-TR acquisitions}

We measured \na \tone using a multi-TR saturation recovery reference experiment that consisted of a series of five FLORET scans with diﬀerent TR = [50, 80, 140, 200, 400] ms and fixed TE = 1 ms. Respective times of acquisition (TA) were [2:24, 3:50, 6:43, 9:36, 19:12] min.

\subsubsection*{Multi-TE acquisitions}

We measured \na \ttwoshort and \ttwolong using a multi-TE reference experiment that consisted of a series of nine FLORET scans with diﬀerent TE = [0.1, 0.5, 1, 2, 5, 10, 20, 50, 90] ms, fixed TR = 200 ms, and TA = 9:36 min for each acquisition.

\subsubsection*{MPRAGE for brain segmentation}

During the \na FLORET scanning session, an additional \hh Magnetization-Prepared Rapid Gradient Echo (MPRAGE) image was acquired with the following parameters: TR = 2300 ms, TE = 2.77 ms, FOV = 240$\times$240 mm\textsuperscript{2}, slice thickness = 1 mm, 176 slices/slab, GRAPPA acceleration factor 2, 1 average, total acquisition time = 3:33 min. 

\subsection*{\na FLORET data reconstruction}

\subsubsection*{Image reconstuction}

For each channel, the raw k-space data was filtered with a Hamming kernel and then reconstructed using standard gridding \cite{fessler2007nufft, pipe2011new} after sampling density compensation \cite{pipe1999sampling}. The images from the 8 channels were combined using image-based coil sensitivity profiles as described by Bydder et al. \cite{bydder2002combination}.

\subsubsection*{Brain segmentation} 

Brain segmentation was processed in Matlab with SPM12 as for the MNF data, but from the MPRAGE data acquired along the \na FLORET data, to generate four ROIs (masks): EYE, CSF, GM and WM. These masks were then applied to the \na FLORET data to quantify mean and standard deviation values of TSC, \tonex, \ttwolong and \ttwoshort for comparison with the MNF data

\subsubsection*{\tone fitting}

For \tone mapping, a mono-exponential fitting was performed voxelwise on the multi-TR FLORET data using the \textit{fit} function in Matlab (Curve Fitting Toolbox) with a nonlinear least square method. The fitting function was $\text{S}_v(\text{TR}) = \alpha(1-e^{-\text{TR}/\text{T}_1}) + \beta$, with $\text{S}_v$ the magnitude of \na signal evolution in voxel $v$ at different TRs, $\alpha$ and $\beta$ fitting variables to adjust data magnitude and baseline offset, respectively. Using the masks from the MPRAGE brain segmentation, \tone ranges were defined as [49-74] ms for EYE+CSF, and [20-48] ms for GM+WM. These ranges times were chosen to coincide with the ranges used for \na MRF dictionary matching. 

\subsubsection*{\ttwoshort and \ttwolong fitting}

For \ttwoshort and \ttwolong mapping, a bi-exponential fitting was performed voxelwise on the multi-TE FLORET data using the \textit{fit} function in Matlab (Curve Fitting Toolbox) with a nonlinear least square method. The fitting function was $\text{S}_v(\text{TE}) = \alpha e^{-\text{TE}/\text{T}_{2,\text{short}}} + (1-\alpha)e^{-\text{TE}/\text{T}_{2,\text{long}}} + \beta$, with $\text{S}_v$ the magnitude of \na signal evolution in voxel $v$ at different TEs, $\alpha$ and $\beta$ fitting variables to adjust data magnitude related to the bi-exponential components and baseline offset, respectively. Using the masks from the MPRAGE brain segmentation, \ttwoshort and \ttwolong were forced to be equal to mono-exponential \ttwo for EYE+CSF (vitreous humor and liquids) in range [49-74] ms. For GM+WM, \ttwoshort and \ttwolong ranges were defined as [0.5-20] and [10-40], respectively. These ranges times were chosen to coincide with the ranges used for \na MRF dictionary matching.

\paragraph{TSC quantification:}

The \na FLORET image with shorter TE = 0.1 ms and longer TR = 200 ms was used to quantify TSC with the same process as for the MNF data: \boneminus correction, normalization, and quantification from the EYE reference. See section "MNF data reconstruction\textbackslash Tissue sodium concentration (TSC)"). 

\subsection*{Statistical analysis}

\subsubsection*{Data types to analyze}

In summary, data types to be analyzed after simultaneous \hhna MRF acquisition and SR post-processing have been performed (where MNF = \hhna MRF + SR) were denominated as follows:

\begin{itemize}
  
    \item 4 datasets: \hh MNF HR (from \hh MRF), \na MNF LR (from \na MRF), \na MNF HR (from \na MRF + SR), and \na FLORET.

    \item 4 ROIs: EYE, CSF, GM, and WM.
    
    \item 7 metrics: PD, \tone and \ttwo from \hh MNF HR, and TSC, \tonex, \ttwoshort and \ttwolong from \na MNF LR, \na MNF HR and \na FLORET.

\end{itemize}

\subsubsection*{Data measurements to analyze}

Measurements from all ROIs, datasets and metrics described above consist of:

\begin{itemize} 

    \item Reference measurements for both PD and TSC quantification were performed 20 times using the N\textsubscript{REF} \% highest voxel values of the reference ROI (EYE) for both metric, with N\textsubscript{REF} = 5\% to 100\% by steps of 5\% (20 N\textsubscript{REF} values). 

    \item ROI measurements of the mean and standard deviation (std) values of all MNF and \na FLORET metrics were performed 20 times in each ROI using the N\textsubscript{ROI} \% highest voxel values of the ROI, with N\textsubscript{ROI} = 5\% to 100\% by steps of 5\% (20 N\textsubscript{ROI} values). The ROI measurements therefore consist of distributions of:

    \begin{itemize}
    
        \item $N$ = 400 \textit{mean} PD and TSC values measured in all ROIs in each subject, from 20 N\textsubscript{ROI}$ \times$ 20 N\textsubscript{REF} values.
        
        \item $N$ = 20 \textit{mean} \hh and \na relaxation times values measured in all ROIs in each subject, from 20 N\textsubscript{ROI} values.
    
    \end{itemize}

\end{itemize}

\subsubsection*{Mean \& standard deviation}

The mean and std values of each distribution of \textit{mean} values for each metric in each ROI were calculated using three statistical methods, and according to the two following steps:

\paragraph{Step 1: \textit{Final} mean $\pm$ std for each subject.} For each metric and each ROI, we measured the final mean and std using the standard statistical method, jackknife resampling and bootstrap resampling methods. Jackknife and bootstrap resampling are usually used to reduce bias and variance in the estimation of statistical parameters (mean and std in our case). 

\begin{enumerate}

    \item \textbf{Standard:} The standard mean $\pm$ std was calculated over all mean values in each dataset (or distribution) of size $N$ = 20 or 400 (depending on the metric), resulting in a \textit{final} \textit{mean} and \textit{std}.     
    
    \item \textbf{Jackknife:} Jackknife resampling \cite{shao2012jackknife, efron1982jackknife} consisted of the following two steps for each subject: 

    \begin{enumerate}

        \item Calculate the mean value of each possible leave-one-out subsample of size ($N-1$) of the full dataset of size $N$ = 20 or 400 mean values (depending on the metric), resulting in a new dataset of $N$ mean values from $N$ subsamples.

        \item Calculate the \textit{final} \textit{mean} and \textit{std} of this new dataset of $N$ mean values from $N$ subsamples. 
        
    \end{enumerate}
                                    
    \item \textbf{Bootstrap:} Bootstrap resampling (or bootstrapping) with replacement \cite{shao2012jackknife, efron1982jackknife} consisted of the following four steps for each subject: 
    
    \begin{enumerate}

        \item Randomly and separately select $N$ values from the full dataset of size $N$ = 20 or 400 mean values (depending on the metric)--which means that some values can be selected multiple times, hence the term "replacement"--to generate a new dataset of $N$ mean values (1 bootstrap).

        \item Calculate the mean value of this new bootstrap dataset of $N$ values.

        \item Repeat steps (a) and (b) $N_{\text{btsp}}$ = 1000 times.

        \item Calculate the \textit{final} \textit{mean} and \textit{std} of the $N_{\text{btsp}}$ mean values for each subject.

    \end{enumerate}
    
\end{enumerate}

\paragraph{Step 2: \textit{Overall} mean $\pm$ std over all subjects.} For each method, the \textit{overall} mean $\pm$ one std of both the \textit{final} \textit{mean} values and the \textit{final} \textit{std} values from each subject was calculated over all subjects.
        
\subsubsection*{Wilcoxon rank-sum test}

The Wilcoxon rank-sum statistical test was applied between the \textit{final} \textit{mean} values from the \textit{standard} method between all metrics from \na MNF HR, LR and FLORET over all subjects in each ROI. Bonferroni correction (BC) was applied to the $p$ values in order to take into account the number of tests ($N_{\text{meas}}$ = 3) of each metric for each tissue: $p_{\text{BC}} = N_{\text{meas}} \cdot p$. Statistically significant differences between measurements, without and with BC, were therefore defined as $p$ and $p_{\text{BC}}$ $\leq$ 0.05.

\section*{Data and Code Availability}

\paragraph{Simultaneous \hhna MRF sequence:} 

The binary files and/or C++ code of the Siemens IDEA sequence (VB17A) and the MATLAB code for image reconstruction are available upon request to the corresponding author, within the Siemens C2P framework   

\paragraph{Super-resolution algorithm:} The MATLAB code of the SR algorithm  \cite{rodriguez2023super} is freely available on Github at the following address: \url{https://github.com/gonggr/Super-Resolution-Sodium-MRI}.

\paragraph{MNF data and statistical analysis:}

The MNF and FLORET brain data, and the MATLAB code used for statistical analysis and for generating all the histograms and boxplots shown in Supplementary Information and in the main article, are freely available on the UltraViolet repository of New York University at the following address: \url{https://doi.org/10.58153/m8qj0-v1936}.

\section*{Ackowledgements}

The authors want to thank Liz Aguilera for recruiting the subjects.

\section*{Author Contributions}

G.M. and M.C. conceived, designed and supervised the study. M.C. and Z.Y. designed and wrote the pulse sequence code for the Siemens MR scanner, and contributed to the design and optimization of the RF pulse train for \hh MRF. L.O., G.G.R. and G.M. contributed to the design and optimization of the RF pulse train for \na MRF. G.G.R. and G.M. contributed to the design and optimization of the super-resolution algorithm, and design of the experiments. G.G.R. carried out the experiments. G.M. wrote the manuscript, and all authors contributed to revisions of the manuscript.  

\section*{Funding}

This project was supported by grant R01EB026456 from the National Institute of Biomedical Imaging and Bioengineering (NIBIB) at National Institutes of Health (NIH), and was also performed under the rubric of the Center for Advanced Imaging Innovation and Research (CAI2R), a NIBIB Biomedical Technology Resource Center (P41EB017183).

\section*{Competing interests} 

The authors declare no competing interests.

\section*{Supplementary Information}

Supplementary information is available for this article in Appendix.

\vfill

\bibliography{references_mnf}

\newpage


\onecolumn

\begin{sidewaystable}[ht]
    \tiny
    \caption{\textbf{Summary of the mean value measurements from \hh MNF HR, \na MNF HR and \na MNF LR in brain at 7 T:} Results are presented as the overall mean $\pm$ one standard deviation (std) of the final mean and std values measured in each ROI and in each subject, from 3 statistical methods: (1) Standard mean $\pm$ std; (2) Jackknife; (3) Bootstrap (n = 1000 bootstraps with replacement). For \na MNF HR and \na MNF LR measurements in the eyes (vitreous humor) and CSF, T\textsubscript{2,long} = T\textsubscript{2,short} $\equiv$ T\textsubscript{2}. See Tables S1, S2 and S3 in Supplementary Information for a detailed summary of all the mean measurements in each subject.}
    \label{tab_summary_mean_values_mnf_brain}
    \centering
    \begin{tabular*}{1\textwidth}{@{\extracolsep\fill}lllllllllllll}
    \toprule
    Statistical Method & Standard & Jackknife & Bootstrap & Standard & Jackknife & Bootstrap & Standard & Jackknife & Bootstrap & Standard & Jackknife & Bootstrap \\
    \cmidrule{1-1}\cmidrule{2-4}\cmidrule{5-7}\cmidrule{8-10}\cmidrule{11-13}
    \textbf{\hh MNF HR} & \multicolumn{3}{@{}l@{}}{~\quad \textbf{PD}} & \multicolumn{3}{@{}l@{}}{~\quad \textbf{T\textsubscript{1} [ms]}} & \multicolumn{3}{@{}l@{}}{~\quad \textbf{T\textsubscript{2} [ms]}} & \multicolumn{3}{@{}l@{}}{~} \\
    \cmidrule{1-1}\cmidrule{2-4}\cmidrule{5-7}\cmidrule{8-10}
    Eye (Vitreous Humor) \\
    \quad\quad Mean $\pm$ Std of Mean & 0.989 $\pm$ 0.003 & 0.989 $\pm$ 0.003 & 0.989 $\pm$ 0.003 & 3913.89 $\pm$ 273.63 & 3913.89 $\pm$ 273.63 & 3913.89 $\pm$ 273.65 & 122.87 $\pm$ 31.25 & 122.87 $\pm$ 31.25 & 122.90 $\pm$ 31.29 \\
    \quad\quad Mean $\pm$ Std of Std & 0.130 $\pm$ 0.022 & 0.000 $\pm$ 0.000 & 0.006 $\pm$ 0.001 & 230.16 $\pm$ 46.11 & 0.58 $\pm$ 0.12 & 11.30 $\pm$ 2.24 & 31.88 $\pm$ 8.72 & 0.08 $\pm$ 0.02 & 1.61 $\pm$ 0.43 \\
    Cerebrospinal Fluid (CSF) \\
    \quad\quad Mean $\pm$ Std of Mean & 0.821 $\pm$ 0.086 & 0.821 $\pm$ 0.086 & 0.821 $\pm$ 0.086 & 3296.38 $\pm$ 145.47 & 3296.38 $\pm$ 145.47 & 3296.37 $\pm$ 145.58 & 145.40 $\pm$ 17.12 & 145.40 $\pm$ 17.12 & 145.35 $\pm$ 17.11 \\
    \quad\quad Mean $\pm$ Std of Std & 0.118 $\pm$ 0.017 & 0.000 $\pm$ 0.000 & 0.006 $\pm$ 0.001 & 390.62 $\pm$ 60.17 & 0.98 $\pm$ 0.15 & 19.67 $\pm$ 2.95 & 44.52 $\pm$ 5.78 & 0.11 $\pm$ 0.01 & 2.26 $\pm$ 0.31 \\
    Gray Matter (GM)  \\
    \quad\quad Mean $\pm$ Std of Mean & 0.739 $\pm$ 0.080 & 0.739 $\pm$ 0.080 & 0.739 $\pm$ 0.080 & 1688.70 $\pm$ 53.25 & 1688.70 $\pm$ 53.25 & 1688.73 $\pm$ 53.40 & 55.25 $\pm$ 3.58 & 55.25 $\pm$ 3.58 & 55.42 $\pm$ 3.37 \\
    \quad\quad Mean $\pm$ Std of Std & 0.083 $\pm$ 0.006 & 0.000 $\pm$ 0.000 & 0.004 $\pm$ 0.000 & 107.93 $\pm$ 5.80 & 0.27 $\pm$ 0.02 & 5.42 $\pm$ 0.37 & 10.84 $\pm$ 2.92 & 0.03 $\pm$ 0.01 & 0.55 $\pm$ 0.15 \\
    White Matter (WM) \\
    \quad\quad Mean $\pm$ Std of Mean & 0.548 $\pm$ 0.062 & 0.548 $\pm$ 0.062 & 0.548 $\pm$ 0.062 & 1027.53 $\pm$ 12.52 & 1027.53 $\pm$ 12.52 & 1027.53 $\pm$ 12.45 & 38.28 $\pm$ 2.97 & 38.28 $\pm$ 2.97 & 38.27 $\pm$ 2.98 \\
    \quad\quad Mean $\pm$ Std of Std & 0.056 $\pm$ 0.005 & 0.000 $\pm$ 0.000 & 0.003 $\pm$ 0.000 & 46.66 $\pm$ 4.50 & 0.12 $\pm$ 0.01 & 2.32 $\pm$ 0.23 & 4.43 $\pm$ 2.31 & 0.01 $\pm$ 0.01 & 0.22 $\pm$ 0.12 \\
    \cmidrule{1-1}\cmidrule{2-4}\cmidrule{5-7}\cmidrule{8-10}\cmidrule{11-13}
    \textbf{\na MNF HR} & \multicolumn{3}{@{}l@{}}{~\quad \textbf{TSC [mM]}} & \multicolumn{3}{@{}l@{}}{~\quad \textbf{T\textsubscript{1} [ms]}} & \multicolumn{3}{@{}l@{}}{~\quad \textbf{T\textsubscript{2,short} [ms]}} & \multicolumn{3}{@{}l@{}}{~\quad \textbf{T\textsubscript{2,long} [ms]}} \\
    \cmidrule{1-1}\cmidrule{2-4}\cmidrule{5-7}\cmidrule{8-10}\cmidrule{11-13}
    Eye (Vitreous Humor) \\
    \quad\quad Mean $\pm$ Std of Mean & 144.37 $\pm$ 0.78 & 144.37 $\pm$ 0.78 & 144.36 $\pm$ 0.78 & 77.00 $\pm$ 11.52 & 77.00 $\pm$ 11.52 & 77.00 $\pm$ 11.52 & 58.38 $\pm$ 8.51 & 58.38 $\pm$ 8.51 & 58.38 $\pm$ 8.51 & 58.38 $\pm$ 8.51 & 58.38 $\pm$ 8.51 & 58.38 $\pm$ 8.51 \\
    \quad\quad Mean $\pm$ Std of Std & 25.51 $\pm$ 4.65 & 0.06 $\pm$ 0.01 & 1.287 $\pm$ 0.26 & 7.96 $\pm$ 1.42 & 0.02 $\pm$ 0.00 & 0.40 $\pm$ 0.08 & 6.11 $\pm$ 1.08 & 0.02 $\pm$ 0.00 & 0.30 $\pm$ 0.05 & 6.11 $\pm$ 1.08 & 0.02 $\pm$ 0.00 & 0.31 $\pm$ 0.06 \\
    Cerebrospinal Fluid (CSF) \\
    \quad\quad Mean $\pm$ Std of Mean & 136.56 $\pm$ 24.12 & 136.56 $\pm$ 24.12 & 136.56 $\pm$ 24.12 & 76.54 $\pm$ 7.70 & 76.54 $\pm$ 7.70 & 76.54 $\pm$ 7.71 & 58.64 $\pm$ 6.19 & 58.64 $\pm$ 6.19 & 58.65 $\pm$ 6.20 & 58.64 $\pm$ 6.20 & 58.64 $\pm$ 6.19 & 58.65 $\pm$ 6.20 \\
    \quad\quad Mean $\pm$ Std of Std & 24.23 $\pm$ 4.18 & 0.06 $\pm$ 0.01 & 1.19 $\pm$ 0.19 & 9.05 $\pm$ 1.04 & 0.02 $\pm$ 0.00 & 0.46 $\pm$ 0.05 & 7.36 $\pm$ 0.72 & 0.02 $\pm$ 0.00 & 0.37 $\pm$ 0.04 & 7.36 $\pm$ 0.72 & 0.02 $\pm$ 0.00 & 0.37 $\pm$ 0.04 \\
    Gray Matter (GM)  \\
    \quad\quad Mean $\pm$ Std of Mean & 89.91 $\pm$ 16.73 & 89.91 $\pm$ 16.73 & 89.92 $\pm$ 16.72 & 43.86 $\pm$ 4.21 & 43.86 $\pm$ 4.21 & 43.86 $\pm$ 4.20 & 11.58 $\pm$ 4.63 & 11.58 $\pm$ 4.63 & 11.58 $\pm$ 4.63 & 27.86 $\pm$ 6.50 & 27.86 $\pm$ 6.50 & 27.87 $\pm$ 6.50 \\
    \quad\quad Mean $\pm$ Std of Std & 14.00 $\pm$ 2.83 & 0.04 $\pm$ 0.01 & 0.70 $\pm$ 0.14 & 3.10 $\pm$ 0.34 & 0.01 $\pm$ 0.00 & 0.16 $\pm$ 0.02 & 2.60 $\pm$ 0.27 & 0.01 $\pm$ 0.00 & 0.13 $\pm$ 0.01 & 2.28 $\pm$ 0.34 & 0.01 $\pm$ 0.00 & 0.11 $\pm$ 0.02 \\
    White Matter (WM) \\
    \quad\quad Mean $\pm$ Std of Mean & 60.67 $\pm$ 13.05 & 60.67 $\pm$ 13.05 & 60.67 $\pm$ 13.05 & 31.26 $\pm$ 3.01 & 31.26 $\pm$ 3.01 & 31.26 $\pm$ 3.01 & 8.67 $\pm$ 2.54 & 8.67 $\pm$ 2.54 & 8.67 $\pm$ 2.54 & 20.21 $\pm$ 4.25 & 20.21 $\pm$ 4.25 & 20.21 $\pm$ 4.25 \\
    \quad\quad Mean $\pm$ Std of Std & 8.54 $\pm$ 2.12 & 0.02 $\pm$ 0.01 & 0.43 $\pm$ 0.11 & 1.35 $\pm$ 0.12 & 0.00 $\pm$ 0.00 & 0.07 $\pm$ 0.01 & 1.59 $\pm$ 0.32 & 0.00 $\pm$ 0.00 & 0.08 $\pm$ 0.02 & 1.17 $\pm$ 0.09 & 0.00 $\pm$ 0.00 & 0.06 $\pm$ 0.00 \\
    \cmidrule{1-1}\cmidrule{2-4}\cmidrule{5-7}\cmidrule{8-10}\cmidrule{11-13}
    \textbf{\na MNF LR} & \multicolumn{3}{@{}l@{}}{~\quad \textbf{TSC [mM]}} & \multicolumn{3}{@{}l@{}}{~\quad \textbf{T\textsubscript{1} [ms]}} & \multicolumn{3}{@{}l@{}}{~\quad \textbf{T\textsubscript{2,short} [ms]}} & \multicolumn{3}{@{}l@{}}{~\quad \textbf{T\textsubscript{2,long} [ms]}} \\
    \cmidrule{1-1}\cmidrule{2-4}\cmidrule{5-7}\cmidrule{8-10}\cmidrule{11-13}
    Eye (Vitreous Humor) \\
    \quad\quad Mean $\pm$ Std of Mean & 144.03 $\pm$ 1.03 & 144.03 $\pm$ 1.03 & 144.04 $\pm$ 1.04 & 60.22 $\pm$ 5.82 & 60.22 $\pm$ 5.82 & 60.22 $\pm$ 5.82 & 50.34 $\pm$ 5.37 & 50.34 $\pm$ 5.37 & 50.34 $\pm$ 5.37 & 50.34 $\pm$ 5.37 & 50.34 $\pm$ 5.37 & 50.34 $\pm$ 5.37 \\
    \quad\quad Mean $\pm$ Std of Std & 23.11 $\pm$ 6.33 & 0.06 $\pm$ 0.02 & 1.15 $\pm$ 0.32 & 2.03 $\pm$ 1.28 & 0.01 $\pm$ 0.00 & 0.10 $\pm$ 0.06 & 1.62 $\pm$ 1.57 & 0.00 $\pm$ 0.00 & 0.08 $\pm$ 0.08 & 1.62 $\pm$ 1.57 & 0.00 $\pm$ 0.00 & 0.08 $\pm$ 0.08 \\
    Cerebrospinal Fluid (CSF) \\
    \quad\quad Mean $\pm$ Std of Mean & 105.21 $\pm$ 29.81 & 105.21 $\pm$ 29.81 & 105.21 $\pm$ 29.82 & 62.04 $\pm$ 6.32 & 62.04 $\pm$ 6.32 & 62.04 $\pm$ 6.32 & 51.86 $\pm$ 6.14 & 51.86 $\pm$ 6.14 & 51.86 $\pm$ 6.14 & 51.86 $\pm$ 6.14 & 51.86 $\pm$ 6.14 & 51.86 $\pm$ 6.14 \\
    \quad\quad Mean $\pm$ Std of Std & 20.19 $\pm$ 7.76 & 0.05 $\pm$ 0.02 & 1.01 $\pm$ 0.39 & 2.33 $\pm$ 0.26 & 0.01 $\pm$ 0.00 & 0.12 $\pm$ 0.02 & 1.86 $\pm$ 0.66 & 0.01 $\pm$ 0.00 & 0.09 $\pm$ 0.03 & 1.86 $\pm$ 0.66 & 0.01 $\pm$ 0.00 & 0.09 $\pm$ 0.03 \\
    Gray Matter (GM)  \\
    \quad\quad Mean $\pm$ Std of Mean & 62.01 $\pm$ 10.22 & 62.01 $\pm$ 10.22 & 62.02 $\pm$ 10.22 & 37.93 $\pm$ 4.61 & 37.93 $\pm$ 4.61 & 37.93 $\pm$ 4.61 & 8.49 $\pm$ 4.71 & 8.49 $\pm$ 4.71 & 8.49 $\pm$ 4.71 & 25.32 $\pm$ 6.83 & 25.32 $\pm$ 6.83 & 25.32 $\pm$ 6.83 \\
    \quad\quad Mean $\pm$ Std of Std & 10.95 $\pm$ 1.33 & 0.03 $\pm$ 0.00 & 0.55 $\pm$ 0.07 & 1.57 $\pm$ 0.06 & 0.00 $\pm$ 0.00 & 0.08 $\pm$ 0.00 & 2.74 $\pm$ 0.19 & 0.01 $\pm$ 0.00 & 0.14 $\pm$ 0.01 & 2.30 $\pm$ 0.12 & 0.01 $\pm$ 0.00 & 0.11 $\pm$ 0.01 \\
    White Matter (WM) \\
    \quad\quad Mean $\pm$ Std of Mean & 51.45 $\pm$ 10.78 & 51.45 $\pm$ 10.78 & 51.46 $\pm$ 10.78 & 38.17 $\pm$ 4.21 & 38.17 $\pm$ 4.21 & 38.17 $\pm$ 4.20 & 9.63 $\pm$ 4.14 & 9.63 $\pm$ 4.14 & 9.63 $\pm$ 4.14 & 25.74 $\pm$ 6.17 & 25.74 $\pm$ 6.17 & 25.74 $\pm$ 6.17 \\
    \quad\quad Mean $\pm$ Std of Std & 9.16 $\pm$ 1.35 & 0.02 $\pm$ 0.00 & 0.46 $\pm$ 0.07 & 1.81 $\pm$ 0.18 & 0.00 $\pm$ 0.00 & 0.09 $\pm$ 0.01 & 3.17 $\pm$ 0.35 & 0.01 $\pm$ 0.00 & 0.16 $\pm$ 0.02 & 2.63 $\pm$ 0.21 & 0.01 $\pm$ 0.00 & 0.13 $\pm$ 0.01 \\
    \bottomrule
    \end{tabular*}
\end{sidewaystable}

\clearpage

\begin{sidewaystable}[th]
    \tiny
    \caption{\textbf{Mean \hh relaxation times and proton density (PD) measured in vivo in brain at 7 T:} Comparison of the overall standard mean values from \hh HR MNF (see Methods, Table \ref{tab_summary_mean_values_mnf_brain} and Table S1 in Supplementary Information) with values from the literature. Results are presented as mean $\pm$ one standard deviation (std).}
    \label{tab_relaxation_1h_comparison}
    \centering
    \begin{tabular*}{1\textwidth}{@{\extracolsep\fill}llllllllllllll}
    \toprule
    \multicolumn{2}{@{}l@{}}{\textbf{Hydrogen (\hh)}} & \multicolumn{3}{@{}l@{}}{~~~ \textbf{Eye (vitreous humor)}} & \multicolumn{3}{@{}l@{}}{~~~ \textbf{Cerebrospinal Fluid (CSF)}} & \multicolumn{3}{@{}l@{}}{~~~ \textbf{Gray Matter (GM)}} & \multicolumn{3}{@{}l@{}}{~~~ \textbf{White Matter (WM)}} \\
    \cmidrule{1-2}\cmidrule{3-5}\cmidrule{6-8}\cmidrule{9-11}\cmidrule{12-14}
    Method & Reference & PD & T\textsubscript{1} [ms] & T\textsubscript{2} [ms] & PD & T\textsubscript{1} [ms] & T\textsubscript{2} [ms] & PD & T\textsubscript{1} [ms] & T\textsubscript{2} [ms] & PD & T\textsubscript{1} [ms] & T\textsubscript{2} [ms] \\
    \midrule
    N/A & \cite{chau2014osteogenesis} & 0.98 $\pm$ 0.00 & ~ & ~ & ~ & ~ & ~ & ~ & ~ & ~ & ~ & ~ & ~ \\
    N/A & \cite{forrester2015eye} & 0.98 $\pm$ 0.00 & ~ & ~ & ~ & ~ & ~ & ~ & ~ & ~ & ~ & ~ & ~ \\
    MRF (Full) & \cite{koolstra2019cartesian} & ~ & 3599 $\pm$ 334 & 145 $\pm$ 12 & ~ & ~ & ~ & ~ & ~ & ~ & ~ & ~ & ~ \\
    IR-TFE & \cite{richdale20097} & ~ & 4250 $\pm$ 0 & ~ & ~ & ~ & ~ & ~ & ~ & ~ & ~ & ~ & ~ \\
    FFE & \cite{richdale20097} & ~ & 5000 $\pm$ 0 & ~ & ~ & ~ & ~ & ~ & ~ & ~ & ~ & ~ & ~ \\
    MRF & \cite{yu2020simultaneous} & ~ & ~ & ~ & ~ & ~ & ~ & ~ & 1702 $\pm$ 187 & 33 $\pm$ 5 & ~ & 1091 $\pm$ 45 & 25 $\pm$ 2 \\
    MRF & \cite{yu2022simultaneous} & ~ & ~ & ~ & ~ & ~ & ~ & ~ & 1451 $\pm$ 160 & 39 $\pm$ 13 & ~ & 962 $\pm$ 92 & 31 $\pm$ 5 \\
    MRF & \cite{rodriguez2022repeatability} & ~ & ~ & ~ & 1.00 $\pm$ 0.36 & 2570 $\pm$ 170 & 102 $\pm$ 19 & 0.87 $\pm$ 0.18 & 1450 $\pm$ 40 & 40 $\pm$ 2 & 0.66 $\pm$ 0.12 & 940 $\pm$ 20 & 32 $\pm$ 1 \\
    PURR & \cite{rooney2007magnetic} & ~ & ~ & ~ & ~ & 4425 $\pm$ 137 & ~ & ~ & 2132 $\pm$ 103 & ~ & ~ & 1220 $\pm$ 36 & ~ \\
    MP2RAGE & \cite{marques2010mp2rage} & ~ & ~ & ~ & ~ & ~ & ~ & ~ & 1920 $\pm$ 160 & ~ & ~ & 1150 $\pm$ 60 & ~ \\
    MPRAGE & \cite{wright2008water} & ~ & ~ & ~ & ~ & ~ & ~ & ~ & 1939 $\pm$ 149 & ~ & ~ & 1126 $\pm$ 97 & ~ \\
    MP2RAGEME & \cite{caan2019mp2rageme} & ~ & ~ & ~ & ~ & ~ & ~ & ~ & 2000 $\pm$ 80 & 32 $\pm$ 1 & ~ & 1190 $\pm$ 20 & 27 $\pm$ 1 \\
    MP2RAGE & \cite{caan2019mp2rageme} & ~ & ~ & ~ & ~ & ~ & ~ & ~ & 1950 $\pm$ 40 & 33 $\pm$ 2 & ~ & 1220 $\pm$ 10 & 27 $\pm$ 1 \\
    2D VFA FLASH & \cite{dieringer2014rapid} & ~ & ~ & ~ & ~ & ~ & ~ & ~ & 2524 $\pm$ 137 & ~ & ~ & 1855 $\pm$ 141 & ~ \\
    IR-SE & \cite{dieringer2014rapid} & ~ & ~ & ~ & ~ & ~ & ~ & ~ & 2065 $\pm$ 69 & ~ & ~ & 1284 $\pm$ 22 & ~\\
    VFA MSE & \cite{leroi2020simultaneous} & ~ & ~ & ~ & ~ & ~ & ~ & 0.67 $\pm$ 0.02 & 2167 $\pm$ 165 & 34 $\pm$ 3 & 0.58 $\pm$ 0.01 & 1400 $\pm$ 51 & 38 $\pm$ 5 \\
    QuICS & \cite{leroi2020simultaneous} & ~ & ~ & ~ & ~ & ~ & ~ & 0.68 $\pm$ 0.01 & 2577 $\pm$ 76 & 49 $\pm$ 4 & 0.59 $\pm$ 0.01 & 1698 $\pm$ 65 & 34 $\pm$ 1 \\
    ME-TSE & \cite{emmerich2019rapid} & ~ & ~ & ~ & ~ & ~ & ~ & ~ & ~  & 55 $\pm$ 2 & ~ & ~ & 39 $\pm$ 5 \\
    Dual SPGR & \cite{sabati2013fast} & ~ & ~ & ~ & ~ & ~ & ~ & 0.80 $\pm$ 0.02 & ~ & ~ & 0.70 $\pm$ 0.02 & ~ & ~ \\
    QUTE (average) & \cite{neeb2008fast} & ~ & ~ & ~ & ~ & ~ & ~ & 0.83 $\pm$ 0.01 & ~ & ~ & 0.67 $\pm$ 0.01 & ~ & ~ \\
    GRE PD+T1 (average) & \cite{abbas2015quantitative} & ~ & ~ & ~ & 0.99 $\pm$ 0.01 & ~ & ~ & 0.83 $\pm$ 0.01 & ~ & ~ & 0.70 $\pm$ 0.01 & ~ & ~ \\
    VFA FLASH & \cite{volz2012correction} & ~ & ~ & ~ & ~ & ~ & ~ & 0.81 $\pm$ 0.01 & ~ & ~ & 0.70 $\pm$ 0.20 & ~ & ~ \\
    %
    %
    SPGR & \cite{mezer2016evaluating} & ~ & ~ & ~ & ~ & ~ & ~ & 0.83 $\pm$ 0.06 & ~ & ~ & 0.74 $\pm$ 0.07 & ~ & ~ \\
    Review (average) & \cite{tofts2003pd} & ~ & ~ & ~ & ~ & ~ & ~ & 0.82 $\pm$ 0.02 & ~ & ~ & 0.70 $\pm$ 0.02 & ~ & ~ \\
    GRE Intermediate TR & \cite{shah2022novel} & ~ & ~ & ~ & 0.99 $\pm$ 0.04 & ~ & ~ & 0.81 $\pm$ 0.06 & ~ & ~ & 0.71 $\pm$ 0.04 & ~ & ~ \\
    GRE Long TR & \cite{shah2022novel} & ~ & ~ & ~ & 0.96 $\pm$ 0.05 & ~ & ~ & 0.82 $\pm$ 0.06 & ~ & ~ & 0.71 $\pm$ 0.04 &~ & ~ \\
    \midrule
    Mean $\pm$ Std & (literature) & 0.98 $\pm$ 0.00 & 4283 $\pm$ 701 & 145 $\pm$ 12 & 0.99 $\pm$ 0.02 & 3498 $\pm$ 1311 & 102 $\pm$ 19 & 0.80 $\pm$ 0.06 & 1990 $\pm$ 352 & 39 $\pm$ 8 & 0.68 $\pm$ 0.05 & 1261 $\pm$ 274 & 32 $\pm$ 5 \\
    Range [min, max] & ~ & [0.98, 0.98] & [3599, 5000] & [145, 145] & [0.96, 1.00] & [2570, 4425] & [102, 102] & [0.67, 0.87] & [1450, 2577] & [32, 55] & [0.58, 0.74] & [940, 1698] & [25, 39] \\
    \midrule
    \hh MNF HR & (this study) & 0.99 $\pm$ 0.00 & 3914 $\pm$ 274 & 123 $\pm$ 31 & 0.82 $\pm$ 0.09 & 3296 $\pm$ 145 & 145 $\pm$ 17 & 0.74 $\pm$ 0.08 & 1689 $\pm$ 53 & 55 $\pm$ 4 & 0.55 $\pm$ 0.06 & 1027 $\pm$ 13 & 38 $\pm$ 3 \\
    \bottomrule
    \end{tabular*}
    \justifying
    \scriptsize{\noindent\textit{Abbreviations:} HR = High Resolution; N/A = Not Available; FFE = Fast Field Echo; IR-TFE = Inversion Recovery Turbo Field Echo; PD = Proton Density; CSF = Cerebrospinal Fluid; PURR = Perturbed Recovery from Inversion; MPRAGE = Magnetization-Prepared Rapid Gradient Echo; MP2RAGE = Magnetization-Prepared 2 Rapid Acquisition Gradient Echoes; MP2RAGEME = MP2RAGE Multi Echo; QuICS = Quantitative Imaging using Configuration States; VFA = Variable Flip Angle; MSE = Multi Spin Echo; ME-TSE = Multi-Echo Turbo Spin Echo; MRF = Magnetic Resonance Fingerprinting; MNF = Multinuclear Fingerprinting; SPGR = Spoiled Gradient Recalled Echo; QUTE = Quantitative T\textsubscript{2}\textsuperscript{*} Image; FLASH = Fast Low Angle Shot; MGE = Multi Gradient Echo; IR-SE = Inversion Recovery Spin Echo. }
\end{sidewaystable}

\clearpage 

\begin{sidewaystable}
    \tiny
    \caption{\textbf{Mean \na relaxation times and total sodium concentration (TSC) measured in vivo in brain at 7 T:} Comparison of the overall standard mean values from \na MNF HR and LR (see Methods, Table \ref{tab_summary_mean_values_mnf_brain} and Tables S2 and S3 in Supplementary Information) with values from the literature. TSC is measured in mM (or mmol/L), with mean TSC in the eyes (vitreous humor) used as internal reference (145 mM). Results are presented as average $\pm$ one standard deviation (std). Since TSC does not depend on the magnetic field, values for comparison were gathered from studies performed at different fields. For eye (vitreous humor) and CSF, T\textsubscript{2,long} = T\textsubscript{2,short} $\equiv$ T\textsubscript{2} (monoexponential relaxation in liquid or quasi-liquid state).}
    \label{tab_relaxation_23na_comparison}
    \begin{tabular*}{1\textheight}{@{\extracolsep\fill}lllllllllllllllll}
    \toprule
    \multicolumn{2}{@{}l@{}}{\textbf{Sodium (\na)}} & \multicolumn{3}{@{}l@{}}{~\quad \textbf{Eye (vitreous humor)}} & \multicolumn{3}{@{}l@{}}{~\quad \textbf{Cerebrospinal Fluid (CSF)}} & \multicolumn{4}{@{}l@{}}{~\quad \textbf{Gray Matter (GM)}} & \multicolumn{4}{@{}l@{}}{~\quad \textbf{White Matter (WM)}} \\
    \cmidrule{1-2}\cmidrule{3-5}\cmidrule{6-8}\cmidrule{9-12}\cmidrule{13-16}
    Method & Reference & TSC [mM] & T\textsubscript{1} [ms] & T\textsubscript{2} [ms] & TSC [mM] & T\textsubscript{1} [ms] & T\textsubscript{2} [ms] & TSC [mM] & T\textsubscript{1} [ms] & T\textsubscript{2,long} [ms] & T\textsubscript{2,short} [ms] & TSC [mM] & T\textsubscript{1} [ms] & T\textsubscript{2,long} [ms] & T\textsubscript{2,short} [ms] \\
    \midrule
    Assay & \cite{kokavec2016biochemical} & 146.7 $\pm$ 3.3 & ~ & ~ & ~ & ~ & ~ & ~ & ~ & ~ & ~ & ~ & ~ & ~ & ~\\
    Used as reference & \cite{winkler1989regional} & 145.0 $\pm$ 0.0 & ~ & ~ & ~ & ~ & ~ & ~ & ~ & ~ & ~ & ~ & ~ & ~ & ~\\
    AWSOS & \cite{mirkes2015high} & 134.0 $\pm$ 6.0 & ~ & ~ & ~ & ~ & ~ & ~ & ~ & ~ & ~ & ~ & ~ & ~ & ~\\
    Mean from references within & \cite{worthoff2019relaxometry} & 139.2 $\pm$ 18.9 & ~ & ~ & ~ & ~ & ~ & ~ & ~ & ~ & ~ & ~ & ~ & ~ & ~\\
    Multi-Echo GRE & \cite{kolodny1993feasibility}\textsuperscript{1} & 143.0 $\pm$ 0.0 & ~ & 46.0 $\pm$ 2.1 & ~ & ~ & ~ & ~ & ~ & ~ & ~ & ~ & ~ & ~ & ~\\
    Multi-Echo GRE & \cite{kohler1989magnetic}\textsuperscript{2} & ~ & ~ & 45.8 $\pm$ 3.6 & ~ & ~ & ~ & ~ & ~ & ~ & ~ & ~ & ~ & ~ & ~\\
    IR + CPMG MRS & \cite{pettegrew1985sodium}\textsuperscript{3} & ~ & 60.9 $\pm$ 0.9 & 60.3 $\pm$ 4.4 & ~ & ~ & ~ & ~ & ~ & ~ & ~ & ~ & ~ & ~ & ~\\
    MRF & \cite{kratzer20213d}\textsuperscript{4} & ~ & ~ & ~ & ~ & 61.9 $\pm$ 2.8 & 46.3 $\pm$ 4.5 & ~ & 35.0 $\pm$ 3.2 & 29.3 $\pm$ 3.8 & 5.5 $\pm$ 1.3 & ~ & 35.0 $\pm$ 3.2 & 29.3 $\pm$ 3.8 & 5.5 $\pm$ 1.3 \\
    MRF & \cite{kratzer2020sodium} & ~ & ~ & ~ & ~ & 67.1 $\pm$ 6.3 & 41.5 $\pm$ 3.4 & ~ & ~ & ~ & ~ & ~ & 38.9 $\pm$ 4.8 & 29.2 $\pm$ 4.9 & 4.7 $\pm$ 1.2 \\
    DA-3DPR & \cite{lommen2018probing} & ~ & ~ & ~ & ~ & ~ & 53.6 $\pm$ 3.9 & ~ & ~ & ~ & ~ & ~ & ~ & 37.7 $\pm$ 2.4 & 5.1 $\pm$ 0.8 \\
    3D GRE & \cite{fleysher2009sodium} & ~ & ~ & ~ & ~ & ~ & 54.0 $\pm$ 4.0 & ~ & ~ & 28.0 $\pm$ 2.0 & ~ & ~ & ~ & 29.0 $\pm$ 2.0 & ~ \\ 
    3D-MERINA & \cite{blunck20183d} & ~ & ~ & ~ & ~ & ~ & 57.2 $\pm$ 6.6 & ~ & ~ & 25.9 $\pm$ 8.3 & 2.0 $\pm$ 1.7 & ~ & ~ & 22.4 $\pm$ 7.8 & 2.0 $\pm$ 2.1 \\
    DA-3DPR & \cite{niesporek2017improved} & ~ & ~ & ~ & ~ & ~ & 46.9 $\pm$ 2.1 & ~ & ~ & 36.4 $\pm$ 3.1 & 5.4 $\pm$ 0.2 & ~ & ~ & 23.3 $\pm$ 2.6 & 3.5 $\pm$ 0.1 \\
    3D SoS & \cite{riemer2018bi} & ~ & ~ & ~ & ~ & ~ & ~ & ~ & ~ & 17.2 $\pm$ 2.0 & 2.9 $\pm$ 0.4 & ~ & ~ & 18.8 $\pm$ 3.2 & 3.1 $\pm$ 0.3 \\
    DA-3DPR & \cite{ridley2018distribution}\textsuperscript{5} & ~ & ~ & ~ & ~ & ~ & ~ & 46.9 $\pm$ 3.0 & ~ & 31.5 $\pm$ 1.8 & 5.0 $\pm$ 0.4 & 38.2 $\pm$ 2.0 & ~ & 38.3 $\pm$ 2.5 & 4.4 $\pm$ 0.3 \\
    Meta-analysis & \cite{ridley2023variability}\textsuperscript{5,6} & ~ & ~ & ~ & ~ & ~ & ~ & 46.5 $\pm$ 8.9 & ~ & ~ & ~ & 37.5 $\pm$ 9.8 & ~ & ~ & ~ \\
    FLORET + Review & \cite{gilles2017multipulse}\textsuperscript{5,7} & ~ & ~ & ~ & ~ & ~ & ~ & 42.2 $\pm$ 12.2 & ~ & ~ & ~ & 42.2 $\pm$ 12.2 & ~ & ~ & ~ \\
    3D Radial & \cite{inglese2010brain}\textsuperscript{5} & ~ & ~ & ~ & ~ & ~ & ~ & 30.5 $\pm$ 3.2 & ~ & ~ & ~ & 19.4 $\pm$ 1.7 & ~ & ~ & ~ \\
    FLORET & \cite{gerhalter2021global}\textsuperscript{5} & ~ & ~ & ~ & ~ & ~ & ~ & 37.9 $\pm$ 1.4 & ~ & ~ & ~ & 31.6 $\pm$ 1.6 & ~ & ~ & ~ \\
    DA-3DPR & \cite{meyer2019cerebral} & ~ & ~ & ~ & 84.6 $\pm$ 6.2 & ~ & ~ & 39.9 $\pm$ 1.6 & ~ & ~ & ~ & 32.8 $\pm$ 2.2 & ~ & ~ & ~ \\
    DA-3DPR & \cite{meyer2019repeatability}\textsuperscript{8} & ~ & ~ & ~ & 103.2 $\pm$ 22.3 & ~ & ~ & 51.5 $\pm$ 6.7 & ~ & ~ & ~ & 40.9 $\pm$ 5.2 & ~ & ~ & ~ \\
    TPI & \cite{bhatia2022quantitative} & ~ & ~ & ~ & 126.6 $\pm$ 11.3 & ~ & ~ & 58.9 $\pm$ 5.7 & ~ & ~ & ~ & 52.5 $\pm$ 4,7 & ~ & ~ & ~ \\
    3D GRE & \cite{petracca2016brain} & ~ & ~ & ~ & ~ & ~ & ~ & 40.3 $\pm$ 3.0 & ~ & ~ & ~ & 27.9 $\pm$ 2.6 & ~ & ~ & ~ \\
    DA-3DPR (Vials) & \cite{adlung2022quantification}\textsuperscript{9} & ~ & ~ & ~ & 156.0 $\pm$ 27.0 & ~ & ~ & 58.0 $\pm$ 8.0 & ~ & ~ & ~ & 53.0 $\pm$ 9.0 & ~ & ~ & ~ \\
    DA-3DPR (CSF) & \cite{adlung2022quantification}\textsuperscript{9} & ~ & ~ & ~ & ~ & ~ & ~ & 61.6 $\pm$ 8.4 & ~ & ~ & ~ & 52.5 $\pm$ 5.1 & ~ & ~ & ~ \\
    DA-3DPR (VH) & \cite{adlung2022quantification}\textsuperscript{9} & ~ & ~ & ~ & ~ & ~ & ~ & 54.6 $\pm$ 4.4 & ~ & ~ & ~ & 46.8 $\pm$ 1.9 & ~ & ~ & ~ \\
    FLORET VFA (old) & \cite{haeger20223t}\textsuperscript{10} & ~ & ~ & ~ & 102.9 $\pm$ 11.4 & ~ & ~ & 47.5 $\pm$ 4.5 & ~ & ~ & ~ & 39.6 $\pm$ 3.8 & ~ & ~ & ~ \\
    FLORET VFA (young)& \cite{haeger20223t}\textsuperscript{10} & ~ & ~ & ~ & 85.7 $\pm$ 10.3 & ~ & ~ & 39.6 $\pm$ 4.4 & ~ & ~ & ~ & 34.5 $\pm$ 5.0 & ~ & ~ & ~ \\
    DA-3DPR & \cite{niesporek2015partial} & ~ & ~ & ~ & 138.0 $\pm$ 4.0 & ~ & ~ & 48.0 $\pm$ 1.0 & ~ & ~ & ~ & 41.0 $\pm$ 3.0 & ~ & ~ & ~ \\
    TPI & \cite{lu2010quantitative} & ~ & ~ & ~ & ~ & ~ & ~ & 38.1 $\pm$ 0.6 & ~ & ~ & ~ & 28.7 $\pm$ 1.2 & ~ & ~ & ~ \\
    TPI & \cite{romanzetti2014mapping}\textsuperscript{6} & ~ & ~ & ~ & ~ & ~ & ~ & ~ & ~ & ~ & ~ & 28.0 $\pm$ 3.7 & ~ & ~ & ~ \\
    \midrule
    Mean $\pm$ Std & (literature) & 141.6 $\pm$ 5.1 & 60.9 $\pm$ 0.9 & 50.7 $\pm$ 8.3 & 113.9 $\pm$ 27.1 & 64.6 $\pm$ 3.7 & 49.9 $\pm$ 5.9 & 46.4 $\pm$ 8.8 & 35.0 $\pm$ 3.2 & 28.1 $\pm$ 6.4 & 4.2 $\pm$ 1.6 & 38.1 $\pm$ 9.6 & 36.9 $\pm$ 2.8 & 28.5 $\pm$ 7.0 & 4.0 $\pm$ 1.2 \\
    Range [min, max] & ~ & [134.0, 146.7] & [60.9, 60.9] & [45.8, 60.3] & [84.6, 156.0] & [61.9, 67.1] & [41.5, 57.2] & [30.5, 61.6] & [35.0, 35.0] & [17.2, 36.4] & [2.0, 5.5] & [19.4, 53.0] & [35.0, 38.9] & [18.8, 38.3] & [2.0, 5.5] \\
    \midrule
    \na MNF HR & (this study) & 144.4 $\pm$ 0.8 & 77.0 $\pm$ 11.5 & 58.4 $\pm$ 8.5 & 136.6 $\pm$ 24.1 & 76.5 $\pm$ 7.7 & 58.6 $\pm$ 6.2 & 89.9 $\pm$ 16.7 & 43.9 $\pm$ 4.2 & 27.9 $\pm$ 6.5 & 11.6 $\pm$ 4.6 & 60.7 $\pm$ 13.1 & 31.3 $\pm$ 3.0 & 20.2 $\pm$ 4.3 & 8.7 $\pm$ 2.5 \\
    \midrule
    \na MNF LR & (this study) & 144.0 $\pm$ 1.0 & 60.2 $\pm$ 5.8 & 50.3 $\pm$ 5.4 & 105.2 $\pm$ 29.8 & 62.0 $\pm$ 6.3 & 51.9 $\pm$ 6.1 & 62.0 $\pm$ 10.2 & 37.9 $\pm$ 4.6 & 25.3 $\pm$ 6.8 & 8.5 $\pm$ 4.7 & 51.4 $\pm$ 10.8 & 38.2 $\pm$ 4.2 & 25.7 $\pm$ 6.2 & 9.6 $\pm$ 4.1 \\
    \bottomrule
    \end{tabular*}
    \scriptsize{\textsuperscript{1}Ex vivo human eyes, 4.7 T.} \\
    \scriptsize{\textsuperscript{2}Ex vivo bovine eyes, 1.9 T.} \\
    \scriptsize{\textsuperscript{3}Ex vivo bovine eyes, 4.7 T.} \\
    \scriptsize{\textsuperscript{4}In this article, gray and white matters were not differenciated and only whole brain tissue was considered (which comprised both gray and white matters together).} \\
    \scriptsize{\textsuperscript{5}Only data from healthy subjects was included.} \\
    \scriptsize{\textsuperscript{6}Meta-analysis over 22 publications.} \\
    \scriptsize{\textsuperscript{7}Average over 15 studies (including the study from the paper), using values measured in parenchyma (GM+WM). The average was calculated over the minimum and maximum TSC values given in Table 5 of the referenced paper, over whole parenchyma (GM+WM).} \\
    \scriptsize{\textsuperscript{8}Std here is the average of the std values measured in this repeatability/reproducibility study (3 scans). } \\
    \scriptsize{\textsuperscript{9}In this study, three quantification methods were compared, using Vials, CSF or vitreous humor (VH) as references. For the vials method, TSC was measured only in normal-appearing GM and WM in 50 stroke patients. For CSF and VH methods, the results in this table are the average of 3 measurements in 3 healthy subjects and the average in 50 stroke patients.} \\
    \scriptsize{\textsuperscript{10}In this study, median TSC was measured in old (67$\pm$9.4 years old) and young (29.2$\pm$6.4 years old) healthy controls.} \\
    \scriptsize{\textit{Abbreviations:} TSC = Total Sodium Concentration; CSF = Cerebrospinal Fluid; GRE = Gradient-Recalled Echo; DA-3DPR = Density Adapted 3-Dimensional Projection Reconstruction; MRF = Magnetic Resonance Fingerprinting; MNF = Multinuclear Fingerprinting; MRS = Magnetic Resonance Spectroscopy; IR = Inversion Recovery; CPMG = Carr-Purcell-Meiboom-Gill; 3D-MERINA = 3D Multi-Echo Radial Imaging; TPI = Twisted Projection Imaging; flexTPI = Flexible TPI; FLORET = Fermat Looped Orthogonally Encoded Trajectories; SoS = Stack-of-Stars; SISTINA = Simultaneous Single-Quantum and Triple-Quantum-Filtered MRI of \na; VH = Vitreous Humor (eye); VFA = Variable FLip Angle.}
\end{sidewaystable}

\clearpage


\includepdf[pages=-]{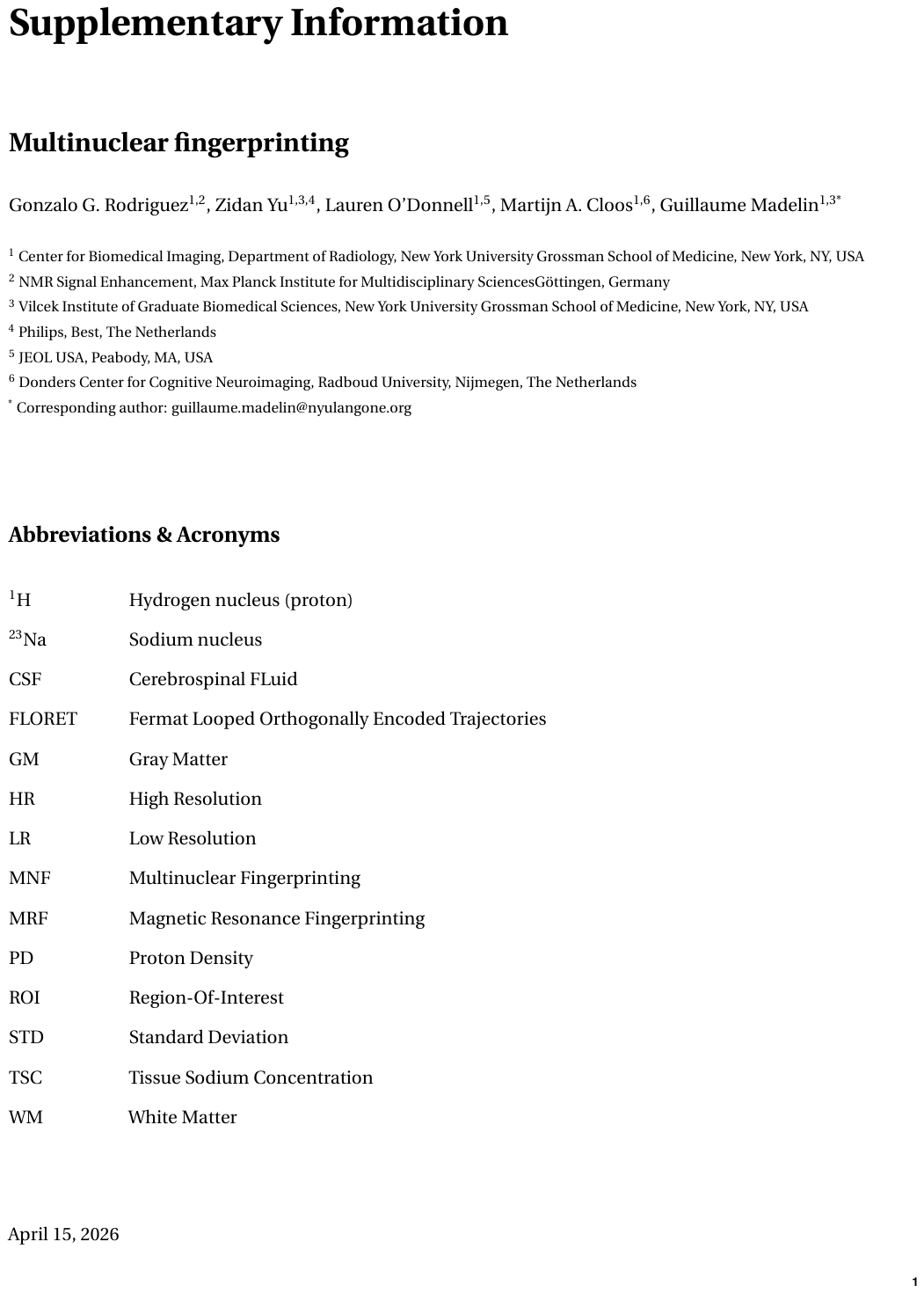}

\end{document}

%% file: abbreviations.tex
\newcommand{\boneplus}{B\textsubscript{1}\textsuperscript{+} }
\newcommand{\boneplusx}{B\textsubscript{1}\textsuperscript{+}}

\newcommand{\boneminus}{B\textsubscript{1}\textsuperscript{-} }
\newcommand{\boneminusx}{B\textsubscript{1}\textsuperscript{-}}

\newcommand{\bzero}{B\textsubscript{0} }

\newcommand{\deltaboneplus}{$\Delta \text{B}_1^+$ }

\newcommand{\deltafzero}{$\Delta \text{f}_0$ }
\newcommand{\deltafzerox}{$\Delta \text{f}_0$}

\newcommand{\na}{\textsuperscript{23}Na }
\newcommand{\nax}{\textsuperscript{23}Na}

\newcommand{\hh}{\textsuperscript{1}H }
\newcommand{\hhx}{\textsuperscript{1}H}

\newcommand{\hhna}{\textsuperscript{1}H/\textsuperscript{23}Na }

\newcommand{\tone}{T\textsubscript{1} }
\newcommand{\tonex}{T\textsubscript{1}}

\newcommand{\ttwo}{T\textsubscript{2} }
\newcommand{\ttwox}{T\textsubscript{2}}

\newcommand{\ttwoshort}{T\textsubscript{2,short} }
\newcommand{\ttwoshortx}{T\textsubscript{2,short}}

\newcommand{\ttwolong}{T\textsubscript{2,long} }
\newcommand{\ttwolongx}{T\textsubscript{2,long}}

\newcommand{\naplus}{Na\textsuperscript{+} }
\newcommand{\naplusx}{Na\textsuperscript{+}}

\newcommand{\nakx}{Na\textsuperscript{+}/K\textsuperscript{+}}